
\input phyzzx
\normalspace


\def\delx{\partial_x}
\def\dely{\partial_y}
\def\delz{\partial_z}
\def\delq{\partial_q}
\def\delqp{\partial_{q'}}

\def\half{{1 \over 2}}
\def\pv{{\int\hskip-12pt-\,}}

\FRONTPAGE
\line{\hfill CNLS-92-02}
\line{\hfill March 1992}
\bigskip
\title{{\bf
    MARINARI-PARISI AND \\
    SUPERSYMMETRIC COLLECTIVE FIELD THEORY \\
     }}
\bigskip
\centerline{Jo\~ao P. Rodrigues
and Andr\'e J. van Tonder}
\centerline{\it Physics Department and Center for Nonlinear
Studies,}
\centerline{\it University of the Witwatersrand, PO Wits 2050, South
Africa}
\bigskip
\abstract
{
A field theoretic formulation of the
Marinari-Parisi supersymmetric matrix model is established and
shown to be equivalent to a recently proposed supersymmetrization of
the
bosonic collective string field theory.  It also corresponds to a
continuum description of super-Calogero
models. The perturbation theory of the model is developed and,
in this approach,
an infinite sequence of vertices is generated.
A class of potentials is identified for which the spectrum is that of
a massless boson and Majorana fermion.   For the harmonic
oscillator case, the cubic vertices
are obtained in an oscillator basis.
For a rather
general class of potentials
it is argued that
one cannot generate from Marinari-Parisi models a continuum limit
similar to that of the $d=1$ bosonic string.
}

\vfill
\endpage

\chapter{INTRODUCTION}

\noindent
As is well known, the double scaling limit [1] of large-$N$ matrix
models describing sums over discretized random surfaces has provided
much insight into the properties of $d\le 1$ strings.
Of these, the one dimensional string [2] exhibits the richest and most
interesting structure (see [3] for a recent review).

The supersymmetric extension of these models has not been
as straightforward as might have been expected, particularly in so far
as their space-time properties are concerned.

Preliminary investigations seem to indicate that the way to lower
dimensional superstrings is not via supermatrices [4].  From a
continuum point of view, one can compute $d\le 1$ correlation
functions [5, 6], but for $d=1$ one finds [6] that the
requirement of locality for the supercharges reduces the space-time
spectrum of the theory to two bosons.   This is substantiated by a
BRST analysis of the theory [7].

In reference [8] Marinari and Parisi introduced a supersymmetric model
which can be thought of as a discretization of one dimensional
superspace (it is to this aspect that we will refer when denoting
these models as Marinari-Parisi models, and not to the specific
choice of potential in [8]).  This model provides a way to stabilize
$d=0$ bosonic potentials.  This results from a well known mechanism of
dimensional reduction [9] (as would be the case
for a corresponding one dimensional Fokker-Planck system), and indeed
a complete analysis of $d\le 0$ bosonic no-unitary multicritical
points via Marinari-Parisi has been carried out recently [10].
However, since Marinari-Parisi models have an interpretation as sums
over discretized surfaces in one dimensional superspace, it is an
interesting question to investigate if a double scaling limit can be
identified leading to a continuum theory of $d=1$ superstrings.
In this sense these models are often referred to in
the literature as possible models of one dimensional superstrings.

Further related work is indicated in reference [11].

Collective field theory [12], in the form developed by Das and
Jevicki [13], provides a very successful description of the one
dimensional bosonic string.  Jevicki and one of the authors [14] have
recently introduced a supersymmetric extension of the collective field
theory by exploiting the metric structure of the bosonic model.
It was emphasized in reference [14]  that the theory that one obtains,
provides, at the classical level, a continuum description of
super-Calogero models of the type introduced by Freedman and Mende
[15]. The relationship between the eigenvalue dynamics of single
matrix
models and Calogero models [16]  is not new [17] and has been a
very fruitful one [17, 18].

More recently, Dabholkar [19] has been able to show that
the Marinari-Parisi model, when restricted to a suitable symmetric
Hilbert space, yields a model which can be expressed in terms of
the $N$ eigenvalues of the bosonic matrix and their
superpartners.
It is not difficult to show that this model is of the
super-Calogero
type.

Therefore, as will be shown in chapter 2 of this paper, supersymmetric
collective field theory [14], Marinari-Parisi [8] and super-Calogero
models [15] are essentially the same system.  In other words,
Dabholkar's reduction of the Marinari-Parisi model yields a
super-Calogero system of which the supersymmetric collective field
theory provides a continuum description.

It is the purpose of this paper to study the properties of the
supersymmetric collective field theory in the presence of an arbitrary
potential.
In this way, as follows from our previous discussion,
we obtain a continuum description of super-Calogero models and
develop a field theory of Marinari-Parisi models.

In chapter 3, the perturbative expansion of the model is developed.
As a result of the nontrivial commutation relations amongst continuum
fields established in reference [14], the hamiltonian of the model
acquires an infinite set of vertices, in contrast to the cubic bosonic
collective string field hamiltonian [20].

Assuming that supersymmetry is preserved at the level of the leading
semiclasical configuration, it follows from the super-Calogero
analogue that the spectrum should be supersymmetric.  In chapter 4 we
show how this can be proved in the continuum.  We also describe a
class of potentials for which the spectrum consists of a massless
boson and Majorana fermion.

In chapter 5 we obtain a manifestly supersymmetric oscillator
expansion fot the cubic vertices in the case of the harmonic
superpotential.

Chapter 6 amounts to a no-go theorem for the possibility of generating
$d=1$ superstrings from Marinari-Parisi models with a rather general
class of potentials.

We will first argue that even
if one regards the true statement of
stationarity as one obtained with respect to the original variables of
the theory (\lq\lq master variables\rq\rq) [21],
one cannot introduce
a lagrange multiplier at the level of the stationarity condition
if
one wishes to preserve supersymmetry.
For the one-dimensional bosonic string, and for a suitable choice of
potential, there exists a rescaling of variables
such that the normalization condition on the density of eigenvalues
becomes essentially a free one.  However, in the supersymmetric case
we will show that for potentials that are homogeneous of arbitrary
degree in the cosmological constant one can identify a rescaling in
terms of which the cosmological constant is scaled out while the
normalization condition remains fixed.  This implies that for these
potentials the time of flight remains finite in the scaling limit and
the mechanism
of generation of an infinite Liouville-like dimension is not present.

In chapter 7 we reexamine the bosonic sector cubic oscillator vertices
of chapter 5 and point out that they are different from those of the
bosonic collective string field theory [20].  We show how this
discrepancy can be traced back to a difference between the turning
point regularization prescriptions of chapter 5 and reference [20].
Adjusting the regularization to agree with that of [20], we argue that
a bosonic sector compatible with the bosonic collective field theory
cannot be generated from a corresponding regularization of the
supercharges considered in this article.

Chapter 8 is reserved for conclusions.

A large portion of this work is contained in AvT's PhD thesis
[23].
During the write-up, we received a preprint by Cohn and Dykstra
[24]
in which related issues are discussed.

\bigskip

\chapter{DISCUSSION OF THE MODELS}

\section {Supersymmetric matrix model}

\noindent
We take as our starting point the supersymmetric matrix model of
Marinari and Parisi.
Our analysis
is based on that of Dabholkar's in [19].

Consider a theory of $N\times N$ hermitian matrices on a
$(1,1)$-dimensional superspace with action
$$
  S = \int dt\,d\theta\,d\bar\theta\, \left(\half{\rm Tr}\, \left(
      \bar D \Phi\, D\Phi\right) + \bar W (\Phi)\right).
   \eqn \matrixS
$$
Here $D \equiv {\partial_{\bar\theta} - i\,\theta\,\partial_t}$, and
we can expand $\Phi$ as
$$
  \Phi \equiv  M + \Psi^\dagger \theta +
     \bar\theta\Psi + \bar\theta\theta F,
  \eqno \eq
$$
where $M$ and $F$ are hermitian, and where we have used the
conventions $(\alpha\beta)^* = \beta^*\alpha^*$ and
$(\partial_{\bar\theta})^* = -\partial_\theta$.

The Feynman diagrams of a matrix theory can be topologically
classified according to the genus.  If one takes ${\bar W}$ to be of
the form
$$
  \bar W (\Phi) = {\rm Tr}\left(
    g_2\,\Phi^2 + {{g_3}\over{\sqrt N}}\,\Phi^3 + \cdots
     + {{g_p}\over{N^{p/2 - 1}}}\,\Phi^p\right),
  \eqn \generalW
$$
then by a topological argument [27]
it follows
that a diagram of genus $\Gamma$ carries a factor $N^{2 - 2\Gamma}$
and can therefore naively be interpreted as a
supertriangulation
of the corresponding diagram in the perturbation expansion of a string
theory with coupling constant $1/N^2$.

Upon quantization one finds the hamiltonian
$$
\eqalign{
  H &= \half\,{\rm Tr} \left(P^2 + {{\partial \bar W(
M)}\over{\partial
         M^*}}\,
       {{\partial \bar W( M)}\over{\partial M}}\right)
       + \sum_{ijkl}\, [\Psi^*_{ji}, \Psi_{kl}]\,
         {{\partial \bar W( M)}\over{\partial M^*_{ij}\,\partial
M_{kl}}}, }  \eqno \eq
$$
and supercharges
$$
\eqalign{
  Q         &= \sum_{ij} \Psi^*_{ij}\left( P^*_{ij} - i\,
                {{\partial \bar W( M)}\over{\partial M^*_{ij}}}
\right), \cr
  Q^\dagger &= \sum_{ij} \Psi_{ij}\left( P_{ij} + i\,
                {{\partial \bar W( M)}\over{\partial M_{ij}}} \right),
\cr
}  \eqn \matrixQ
$$
where $[P_{ij}, M_{kl}] = -i\,\delta_{ik}\delta_{jl}$ and
$\{\Psi_{ij},\Psi_{kl}\} = \delta_{ik}\delta_{jl}$.

Now let $U$ be the unitary transformation that diagonalizes the
bosonic component $ M$ of the matrix $\Phi$.  In general the
fermionic components $\Psi$ and $\Psi^\dagger$ of
$\Phi$ will not be diagonalized by $U$.  Nevertheless, writing
the diagonal elements as
$$
  (U\Phi U^\dagger)_{ii}
   \equiv \lambda_i + \bar\theta\,\psi_i + \psi_i^\dagger\,\theta +
   \bar\theta\theta f_i,     \eqno \eq
$$
we shall see that we can define an invariant subspace
consisting of those wave functions that depend
only on $(\lambda, \psi^\dagger)$.  Therefore it will
make sense to restrict the theory to this subspace.

Changing variables from $M_{ij}$ to the $N$ eigenvalues
$\lambda_i$ and the $N(N-1)/2$ angular variables on which $U$ depends,
one gets the decomposition (see Dabholkar [19])
$$
  {\partial\over{\partial M_{ij}}} = \sum_m U^\dagger_{jm}U_{mi}\,
     {\partial\over{\partial \lambda_m}}
     + \sum_{k\ne l}{{U_{ki}U_{jl}^\dagger\,\tilde A_{kl}}\over
         {(\lambda_k - \lambda_l)}},    \eqno \eq
$$
where the angular derivative $\tilde A_{kl}$ is defined by $\tilde
A_{kl} \equiv \sum_m U_{lm} \,\partial/\partial U_{km}$.

In terms of the new variables the supercharges \matrixQ\ become
$$
\eqalign{
  Q        &= \sum_m \hat\Psi^*_{mm}\left(-i\,
               {\partial\over{\partial\lambda_m}} - i\,
                {{\partial \bar W(\lambda)}\over{\partial \lambda_m}}
                \right) -i \sum_{k\ne l}
                {{\hat\Psi^*_{kl}\,\tilde A^*_{kl}}\over{(\lambda_k -
                \lambda_l)}}, \cr
  Q^\dagger &= \sum_m \hat\Psi_{mm}\left(-i\,
               {\partial\over{\partial\lambda_m}} + i\,
                {{\partial \bar W(\lambda)}\over{\partial \lambda_m}}
                \right) -i \sum_{k\ne l}
                {{\hat\Psi_{kl}\,\tilde A_{kl}}\over{(\lambda_k -
                \lambda_l)}}, \cr
}  \eqn \changedQ
$$
where $\hat\Psi \equiv U \Psi U^\dagger$.  Using
$\hat\Psi_{mm} = \psi_m$, one can write a wave function depending
only on $(\lambda, \psi^\dagger)$ as
$$
  \phi (\lambda, \psi^\dagger) \equiv f(\lambda) \prod_k
       \psi^\dagger_{m_k}|0\rangle = f(\lambda) \prod_k
         \hat\Psi^*_{m_k m_k}|0\rangle.
  \eqn \WfInSubspace
$$
With a little patience one can now show that the subspace of
such
wavefunctions is indeed invariant under the action of the supercharges
\changedQ\ and that in the subspace the supercharges reduce to
$$
\eqalign{
  Q &= \sum_i \psi_i^\dagger\left( -i\,{\partial\over{\partial
\lambda_i}} - i\,
       {{\partial \bar W(\lambda)}\over{\partial\lambda_i}}\right),
\cr
  \bar Q &= \sum_i \psi_i\left( -i\,{\partial\over{\partial
\lambda_i}}
             +i\,{{\partial
             \bar W(\lambda)}\over{\partial\lambda_i}}
             - i\sum_{l\ne i}{1\over{\lambda_i - \lambda_l}}
             \right). \cr
}  \eqn \subspaceQ
$$
At first glance these expressions may seem inconsistent in that $\bar
Q$ appears to have lost its property of conjugacy to $Q$.
This paradox is resolved by the
observation
that $Q$ and $\bar Q$ are indeed hermitian conjugates with respect to
the inner
product on the original Hilbert space.  The original inner product
reduces to a nontrivial inner product on the subspace.

To see how this works, let us review the analysis of Jevicki and
Sakita [12], modified here to include fermionic degrees of freedom.
Take $\phi_1$ and $\phi_2$ to be of the form \WfInSubspace.  The inner
product on the original Hilbert space can be written as
$$
\eqalign{
  \langle \phi_1 |\phi_2\rangle
    &=  \int (d M)(d\Psi^\dagger)(d\Psi)\,
           e^{-{\rm Tr}
           (\Psi^\dagger\Psi)}
        \phi_1^*(\lambda,\psi)\,\phi_2(\lambda,\psi^\dagger)
\cr
    &=  \int (d\lambda)(d\hat\Psi^\dagger)
           (d\hat\Psi)\, e^{-{\rm Tr}
           (\hat\Psi^\dagger\hat\Psi)} J (\lambda)\,
        \phi_1^*(\lambda,\psi)\,\phi_2(\lambda,\psi^\dagger)
\cr
    &=  \int (d\lambda)(d\psi^\dagger)(d\psi)\,e^{-\psi^\dagger\psi}
           J (\lambda)\,
        \phi_1^*(\lambda,\psi)\,\phi_2(\lambda,\psi^\dagger),
\cr
}  \eqno \eq
$$
where $dM = \prod_i dM_{ii} \prod_{j>i}dM_{ij}\,d\bar M_{ij}$
integrates over the independent degrees of freedom of a hermitian
matrix only and where we have
used the fact that the fermionic
measure $\exp[-{\rm Tr}(\Psi^\dagger\Psi)]$ is invariant
under
$\Phi\rightarrow U\Phi U^\dagger$ (this measure has to be
included when one writes the fermionic part of the inner product as a
Berezinian integral [30]).

One sees that the inner product in \? differs from
the trivial inner product on the subspace in that one has to
include the measure $ J (\lambda)$. In general, this measure
is given by the integral
over the spurious degrees of freedom of the jacobian associated to
the change of variables.

If we now rescale the wave functions as $\phi\rightarrow{
J}^{1/2}\phi$, we can use the trivial
inner product on the subspace provided we rescale the momenta as $p_i
\rightarrow
 J^{1/2}\,p_i\, J^{-1/2} = p_i + {i\over 2}\, {{\partial(\ln
J)}/{\partial\lambda_i}}$.

Applying this transformation to the supercharges \subspaceQ\ we find
$$
\eqalign{
  Q &\to \sum_i \psi_i^\dagger\left( -i\,{\partial\over{\partial
\lambda_i}}
       + \half\, i\, {{\partial\ln  J}\over{\partial\lambda_i}}
       - i\,
       {{\partial \bar W(\lambda)}\over{\partial\lambda_i}}\right),
\cr
  \bar Q &\to \sum_i \psi_i\left( -i\,{\partial\over{\partial
\lambda_i}}
          + \half\, i \,{{\partial\ln  J}\over{\partial\lambda_i}}
             +i\,{{\partial \bar
             W(\lambda)}\over{\partial\lambda_i}}
             - i\sum_{l\ne i}{1\over{\lambda_i - \lambda_l}}
             \right). \cr
}  \eqno \eq
$$

The most efficient way to solve for $ J$ is simply to use the fact
that now $\bar Q^\dagger = Q$ with respect to the trivial inner
product in the subspace. We find
$$
  {{\partial\ln  J}\over{\partial\lambda_i}}
             = \sum_{l\ne i}{1\over{\lambda_i - \lambda_l}},
  \eqno \eq
$$
and we get an effective theory with trivial inner product and
supercharges
$$
\eqalign{
  Q &= \sum_i \psi_i^\dagger\left(p_i
       - i\,
       {{\partial W
         (\lambda)}\over{\partial\lambda_i}}\right), \cr
  Q^\dagger &= \sum_i \psi_i\left(p_i
       + i\,
       {{\partial W
         (\lambda)}\over{\partial\lambda_i}}\right), \cr
}  \eqn \QeffMP
$$
where the effective potential $W$ is given by
$$
\eqalign{
  W
              &= \bar W - \sum_{l<k}\ln(\lambda_k-\lambda_l).  \cr
}  \eqn \WeffMP
$$

\section {The super-Calogero model}

\noindent
The bosonic Calogero model [16] provides an example of an
exactly
solvable $N$-particle quantum mechanical system.
Its hamiltonian
in the centre of mass frame of reference is given by
$$
  H_B = \half\sum_{i=1}^N \, p_i^2 + V_B(x_1,\ldots,x_N),
  \eqn \bCalogeroA
$$
where the potential $V_B$ is chosen to be
$$
  V_B(x_1,\ldots,x_N) = {{\omega^2}\over 2}\sum_i x_i^2
     + {\varepsilon^2\over 2}\sum_{i\ne j}{1\over{(x_i - x_j)^2}}.
\eqn \bCalogeroB $$

We see that the potential consists of a harmonic piece as well as a
singular term describing
a repulsive interparticle force.  The equivalence with matrix
models follows from the fact (demonstrated above for the
supersymmetric case) that when one quantizes the singlet sector
of a $U(N)$ invariant matrix model, there appears
a jacobian
that can be reinterpreted as an effective repulsion between the
eigenvalues [27].  This effective interaction
is identical to the singular term in \bCalogeroB.
This will be seen more clearly in the supersymmetric case discussed
below.

The supersymmetric generalization of the Calogero model was first
investigated by Freedman and Mende in [15].
For completeness, we repeat their construction here.
Using the approach of Witten [31] one introduces, in addition to the
bosonic coordinates $x_i$, the fermionic coordinates
$\psi_i$ and $\psi_i^\dagger$ satisfying the standard anti-commutation
relations
$\{\psi_i,\psi_j\}=0$, $\{\psi_i^\dagger,\psi_j^\dagger\}=0$ and
$\{\psi_i,\psi_j^\dagger\}=\delta_{ij}$.  One then defines
supercharges in terms of a so-called superpotential
$W(x_i,\ldots,x_N)$ as
$$
 \eqalign{
  Q &\equiv \sum_i \psi_i^\dagger\left(
            p_i - i\, {{\partial W}\over{\partial x_i}}\right),  \cr
  Q^\dagger &\equiv \sum_i \psi_i\left(
            p_i + i\, {{\partial W}\over{\partial x_i}}\right)  \cr
 }  \eqn \sCalogeroA
$$
satisfying $\{Q,Q\} = 0 = \{Q^\dagger,Q^\dagger\}$.  The hamiltonian
is constructed as
$$
 \eqalign{
  H_S &\equiv \half \{  Q, Q^\dagger   \}   \cr
      &=  \half\sum_i \left(p_i^2
             + \left(
            {{\partial W}\over{\partial x_i}} \right)^2 \right)
      +\half\sum_{i,j}\, [\psi^\dagger_i,\psi_j]\,
                {{\partial^2 W}\over{\partial x_i \partial x_j}}
 }  \eqn \sCalogeroB
$$
and commutes with $Q$ and $Q^\dagger$.

The model studied in [15] corresponded to choosing
$$
  W(x_1,\ldots,x_N) = {\omega\over 2}\sum_{i}x_i^2
              + {\varepsilon\over 2}\sum_{i\ne j}\ln |x_i-x_j|,
\eqn \CalogeroW
$$
in which case one finds, after some algebra [15],
$$
 \eqalign{
  H_S &= \half\sum_{i} \, p_i^2 +
          {{\omega^2}\over{2}}\sum_{i}x_i^2
           + {{\varepsilon^2}\over 2}\sum_{i\ne j}{1\over{(x_i -
            x_j)^2}} \cr
      &+ \omega \sum_i \psi_i^\dagger\psi_i
       - {\varepsilon\over 2}\sum_{i \ne j}\,
       [\psi_i^\dagger-\psi_j^\dagger,\psi_i]\,{1\over{(x_i-x_j)^2}}
         \cr
     &- {\omega\over 2}\, (1 - \varepsilon N) (N-1). \cr
 }  \eqn \sCalogeroH
$$
The bosonic part of this supersymmetric hamiltonian
coincides, apart from an additive constant, with the original Calogero
model \bCalogeroA\ and \bCalogeroB.

The superpotential \CalogeroW\ is just a special case
of the effective potential
 \WeffMP\ found to describe the singlet sector of the
Marinari-Parisi model in the previous section.  At this level one
therefore sees a correspondence between the Marinari-Parisi
and super-Calogero models.

\section{Continuum description}

\noindent
We now follow references [14] and
[17]
to set up a continuum approximation of the discrete model
\QeffMP, or equivalently \sCalogeroA\ and \sCalogeroB.
This approximation is assumed to become exact in the limit
$N\to\infty$, where the discrete distribution $x_i$ will
approximate a continuous density.  With this in mind, we introduce
the continuum index $x$ and define the fields
$$
  \phi(x) \equiv \delx\varphi \equiv \sum_i \delta(x-x_i),
\qquad
  \phi\sigma(x)\equiv -\sum_i\delta(x-x_i)\,p_i,
$$
$$
  \psi(x) = -\sum_i \delta(x-x_i)\,\psi_i, \qquad
  \psi^\dagger(x) = -\sum_i \delta(x-x_i)\,\psi^\dagger_i.
  \eqn \DiscToCont
$$
These fields satisfy commutation relations
$$
 \eqalign{
  [\sigma(x), \varphi(y)] &= -i\, \delta (x-y),  \cr
  [\sigma(x), \psi^\dagger(y)] &= i\, {{\psi^\dagger}\over\phi}(x)\,
                          \delx\delta(x-y),   \cr
  [\sigma(x), \psi(y)] &= i\, {{\psi}\over\phi}(x)\,
                          \delx\delta(x-y),   \cr
  \{\psi(x), \psi^\dagger(y)\} &= \phi(x)\,\delta(x-y).   \cr
 } \eqn \cCommutations
$$
These equalities are proved
using the identity
$$\delta(x-x_i)f(x) =
\delta(x-x_i)f(x_i),   \eqno\eq$$
 where $f$ is an arbitrary function.

Let us now rewrite the supercharges \QeffMP\ or \sCalogeroA\ in terms
of the continuum fields.  At the classical level one has $\int dx\,
{{\delta(x-x_i)\,\delta(x-x_j)}/{\phi(x)}} = \delta_{ij}$, which
can be
established via \?.  Also, by the chain rule
$\partial
W/\partial{x_i} = \delta W/\delta\varphi(x_i)$.
Therefore the
supercharges can equivalently be expressed as
$$
 \eqalign{
  Q &= \int dx\,\psi^\dagger(x)\,(\sigma(x) -i\,W_{;x}),  \cr
  Q^\dagger &= \int dx\,\psi(x)\,(\sigma(x) +i\,W_{;x}),  \cr
 } \eqn \continuumQ
$$
where we have used the notation $W_{;x}\equiv\delta W/\delta\varphi(x)$.

In a careful quantum mechanical treatment of bosonic matrix
models, additional terms arise in the corresponding
continuum hamiltonian [12, 35].
However, we wish to establish here the supersymmetrization of the
cubic collective field theory of [13]. For this
model, perturbative calculations of bosonic scattering amplitudes [20]
reproduce exactly those obtained in the continuum approach.  In the
free potential case [26] an exact solution exhibits two
single particle branches, one of which can always be described
semiclassically by the inclusion of the extra terms.
We assume that a similar
mechanism would apply here and
postulate \continuumQ\ to be the full quantum mechanical
supercharges.  The extra terms are easily taken into account by adding
to the supercharges a term proportional to [23] $\int
dx\,\psi^\dagger(x)\,{\phi_{,x}\over\phi}$.

We now construct the continuum hamiltonian from the supercharges
\continuumQ. We find, using $H=\half\{Q,Q^\dagger\}$, that
$$
\eqalign{
  H &= \half\int dx\, \phi\,\sigma^2 +\half\int dx\,
       \phi\,(W_{;x})^2  \cr
    &- \half\int {{dx}\over\phi}\,[\psi^\dagger,\psi]\,\delx W_{;x}
      + \half\int dx \int dy\, [\psi^\dagger(x),\psi(y)]\,W_{;xy}.
}  \eqn \continuumH
$$
{}From the definition of the field $\phi$, we see that the continuum
hamiltonian must be accompanied by the constraint
$$
  \int dx\, \phi(x) = N.
$$

We now have a language in which to compare the
Marinari-Parisi/super-Calogero model with
the supersymmetrization of the bosonic collective
field theory as constructed in [14], which we do in the next
section.

\section {Supersymmetrized collective field
model}

\noindent
In [14] a supersymmetric extension of the bosonic collective field
theory was described.
This was done by noting
that the collective field theory can be seen as a metric
theory, which can be supersymmetrized via
a standard procedure.

To see how this works, observe that the kinetic term of the bosonic
collective lagrangian
$$
  L = \half\int {{dx}\over\phi}\, \dot\varphi^2 -
            {{\pi^2}\over 6}\int dx\,\phi^3 - \int dx\,v\phi
  \eqno \eq
$$
can be written in the form
$$
 L_T = \half\int dx \int
dy\,\dot\varphi(x,t)\,g_{xy}\,\dot\varphi(y,t),
$$
where the continuous index metric is given by
$$
 g_{xy}(\varphi) = {1\over{\phi(x)}} \,\delta (x-y).
 \eqn \gxy
$$

A standard supersymmetrization of a theory of this type is given by
$$
 \eqalign{
  L &= \half (\dot q^a g_{ab}\, \dot q^b - g^{ab}\,\partial_a W\,
        \partial_b W) + i\,\psi^{\dagger a}g_{ab}\,\dot\psi^b   \cr
    &+ i\,\psi^{\dagger a}\Gamma_{bc,a}\,\dot q^c\psi^b
      + \half\,[\psi^{\dagger c},\psi^d]\,\Gamma^a_{cd}\,\partial_a W
      - \half\,[\psi^{\dagger a},\psi^b]\,\partial_a\partial_b W,
 }   \eqno \eq
$$
where $q^a$ are the bosonic variables, $g_{ab}$ is a metric, $W$ is a
superpotential and $\Gamma^c_{ab}$ are the Christoffel symbols.
This can be obtained by a classical point transformation $x^i \equiv
x^i(q^a)$, $\psi^i = (\partial x^i/\partial q^a)\psi^a \equiv
e^i_a\psi^a$ from the lagrangian
$$
  L = \sum_{i=1}^N \left(\half\dot x_i^2 -\half(\partial_i W)^2\right)
    + i\sum_{i=1}^N \,\psi_i^\dagger\dot\psi_i
    -\sum_{ij}\,[\psi^\dagger_i,\psi_j]\,\partial_i\partial_j W,
$$
which is  a multidimensional generalization of one-dimensional
supersymmetric quantum mechanics [31].

Note the expressions of the form $\half\,[\psi^{\dagger},\psi]$ here.
Classically, taking $\psi$ and $\psi^\dagger$ to be Grassman
variables, we have the identity
$\half\,[\psi^{\dagger},\psi] =\psi^\dagger \psi$.  At the
quantum level, however,
this replacement would destroy supersymmetry unless an additional
term were added
to the bosonic potential in \?.  Thus \? is the proper lagrangian to
use for quantization.

In our case the $q^a$ are replaced by $\varphi(x)$ and $g_{xy}$ is
given by \gxy. One obtains the lagrangian
$$
 \eqalign{
  L &= \half \int{{dx} \over \phi}\, {\dot \varphi}^2
      - \half\int dx\,\phi\,(W_{;x})^2 \cr
     &+ {i} \int {{dx} \over \phi}\,\psi^\dagger\dot\psi
      + {i} \int{{dx} \over \phi}\,\dot\varphi\,
         \delx\left({{\psi^\dagger} \over \phi}\right) \psi
           \cr
     &+ \half\int{{dx} \over \phi}\, [\psi^\dagger,\psi] \,\delx
        W_{;x}
      - \half\int dx\int dy\, [\psi^\dagger(x),\psi(y)]\, W_{;xy}. \cr
 }
  \eqn \nonsymmetricL
$$
This is equivalent, via a partial integration, to
$$
 \eqalign{
  L &= \half \int{{dx} \over \phi}\, {\dot \varphi}^2
      - \half\int dx\,\phi\,(W_{;x})^2 \cr
     &+ {i \over 2} \int {{dx} \over \phi}\,(\psi^\dagger\dot\psi
             - {\dot \psi}^\dagger \psi)
      + {i \over 2} \int{{dx} \over \phi}\,\dot\varphi
         \left[\delx\left({{\psi^\dagger} \over \phi}\right) \psi
          -\psi^\dagger \delx \left({\psi \over\phi}\right)\right] \cr
     &+ \half\int{{dx} \over \phi}\, [\psi^\dagger, \psi] \,\delx
         W_{;x}
      - \half\int dx\int dy\, [\psi^\dagger(x),\psi(y)]\, W_{;xy}. \cr
 }
  \eqn \symmetricL
$$

Our motivation for this rewriting is that in the form \symmetricL\ the
momentum conjugate to the field $\varphi$ is manifestly hermitian,
while in \nonsymmetricL\ it is nonhermitian.  Even so, one would like
the respective theories to be equivalent at the quantum level.
Though it would take us too far afield to show it here, this indeed
turns out to be the case:  in both cases we can define a hermitian
$\sigma$ which satisfies the commutation relations \cCommutations\ and
in terms of which the hamiltonian can be written in the form
\continuumH.

The conjugate momenta are given by
$$
     p(x)  = {{\partial L} \over {\partial\dot\varphi}}
           = {{\dot\varphi}\over\phi}
             + {i\over 2} \left[\delx\left({{\psi^\dagger} \over
              \phi}\right) {\psi\over\phi}
              -{{\psi^\dagger}\over\phi} \delx \left({\psi \over
                \phi}\right)\right],    \eqno \eq
$$
$$
     \Pi = {{\partial L} \over {\partial\dot\psi}}
         = {i\over 2} {{\psi^\dagger}\over\phi}, \qquad
     \Pi^\dagger = {{\partial L} \over {\partial\dot\psi^\dagger}}
         = -{i\over 2} {{\psi}\over\phi}. \eqno \eq
$$

Identifying the second class constraints
$$
 \eqalign{
    \chi &= \Pi - {i\over 2} {{\psi^\dagger}\over\phi}, \cr
    \bar\chi &= \Pi^\dagger + {i\over 2} {{\psi}\over\phi},  \cr
 } \eqno \eq
$$
one uses Dirac brackets [14, 29] to obtain the following equal
time brackets
$$
  [\varphi(x),\varphi(y)] = 0, \quad [\varphi(x),p(y)] = i\delta(x-y),
  \quad \{\psi(x),\psi^\dagger(y)\} = \phi(x)\delta(x-y), \eqn
\collCommA
$$
$$
  [\varphi(x),\psi(y)] = [\varphi(x),\psi^\dagger(y)] = 0,
  \eqn \collCommB
$$
$$
  [\,p(x),\psi(y)] = {i\over
   2}{{\psi(y)}\over{\phi(y)}}\,\delx\delta(x-y),
  \quad
  [\,p(x),\psi^\dagger(y)] = {i\over
     2}{{\psi^\dagger(y)}\over{\phi(y)}}\,\delx\delta(x-y). \eqn
\collCommC
$$
In terms of
$$
  \sigma(x) \equiv p(x)
              - {i\over 2} \left[\delx\left({{\psi^\dagger} \over
               \phi}\right) {\psi\over\phi}
               -{\psi^\dagger\over\phi} \delx \left({\psi \over
                 \phi}\right)\right]    \eqn \sigmaDef
$$
the hamiltonian then takes the form
$$
\eqalign{
  H &= \half\int dx\, \sigma\phi\sigma +\half\int dx\,
       \phi\,(W_{;x})^2  \cr
    &- \half\int {{dx}\over\phi}\,[\psi^\dagger,\psi]\,\delx W_{;x}
      + \half\int dx \int dy\, [\psi^\dagger(x),\psi(y)]\,W_{;xy},
}  \eqno \eq
$$
while from equations \collCommA, \collCommC\ and \sigmaDef\ it
follows that
$$
 [\sigma(x),\psi(y)] = i\,{{\psi}\over{\phi}}(x)\,\delx\delta(x-y),
 \qquad
 [\sigma(x),\psi^\dagger(y)] = i\,
    {{\psi^\dagger}\over{\phi}}(x)\,\delx\delta(x-y).
\eqn\SigmaPsiComm $$

The collective field $\phi$ satisfies the constraint
$$
  \int dx\,\phi(x) = N.
$$

This hamiltonian and these commutation relations are
identical with those derived in the continuum description of the
Marinari-Parisi/super-Calogero model in \cCommutations\ and
\continuumH,
thus completing the chain of
equivalences between the Calogero model, the
supersymmetrized collective field theory and the supersymmetric matrix
model.

\chapter {PERTURBATION THEORY}

\noindent
In this chapter we write down the general perturbation theory of
the supersymmetric model given an arbitrary potential.  We show that
in this approach an infinite sequence of
higher order interactions is generated,
in contrast to the bosonic collective field theory, where
there are only cubic interactions.

\section {Perturbative expansion}

As follows from our discussion of the previous section, we are led to
consider general effective superpotentials of the form
$$
  W = \int dx\, \bar W(x)\,\delx\varphi - {1\over 2}
     \int dx \int dy\, \ln |x-y|\,\delx\varphi\,\dely\varphi.
     \eqn \generalContW
$$
For the special case $\bar W(x) = {\omega\over 2}\,x^2$, this is just
a rewriting in terms of the continuum fields \DiscToCont\ of
the harmonic plus repulsive superpotential \CalogeroW\ of the
super-Calogero model.  In general $\bar W$ will depend on $N$ as in
\generalW.

What is unusual about the commutations \SigmaPsiComm\ is the
fact
that the bosonic momentum does not commute with the fermionic fields.
We observe that from \collCommA\ and \collCommC\ we have
$$
 \biggl[ p(x), {{\psi(y)} \over {\sqrt{\phi(y)}}}\biggr] = 0,
 \qquad
 \biggl[ p(x), {{\psi^\dagger(y)} \over {\sqrt{\phi(y)}}}\biggr] = 0.
 \eqno \eq
$$
We therefore rescale $\psi(x)\to\sqrt{\phi(x)}\,\psi(x)$;
$\psi^\dagger(x)\to\sqrt{\phi(x)}\,\psi^\dagger(x)$ and, defining
$\tilde v(x) \equiv \bar W'(x)$, we obtain for the lagrangian
$$
 \eqalign{
  L &= \half \int{{dx} \over \phi}\, {\dot \varphi}^2
      - \half\int dx\,\phi\,(W_{;x})^2 \cr
     &+ {i \over 2} \int {{dx}}\,(\psi^\dagger\dot\psi
             - {\dot \psi}^\dagger \psi)
      + {i \over 2} \int{{dx} \over \phi}\,\dot\varphi
         \left[\delx{{\psi^\dagger}} \psi
          -\psi^\dagger \delx {\psi}\right] \cr
     &+ \half\int{{dx}}\, [\psi^\dagger,\psi] \,\delx W_{;x}
      - \half\int dx\int dy\,
        [\psi^\dagger(x),\psi(y)]\sqrt{\phi(x)}\, W_{;xy}\,
          \sqrt{\phi(y)}, \cr
 }
  \eqn \rescaledL
$$
for the hamiltonian
$$
  \eqalign{
   H &= {1 \over {2}} \int{{dx}\over\phi}\,
        \left(\phi p - {i\over 2} \left[(\delx\psi^\dagger)\psi
              -\psi^\dagger (\delx \psi)\right]\right)^2 \cr
      &+ {{1} \over 2}\int dx\,\phi(x)
            \left(\pv dy\,{{\dely \varphi} \over
             {x-y}} - \tilde v(x) \right)^2 \cr
      &- \half\int dx\, [\psi^\dagger, \psi] \,\,{d\over dx}\left(
              \pv dy\,{{\dely\varphi}\over{x-y}} -
               \tilde v(x)\right)  \cr
     &+ \half\int dx\,\biggl[\psi^\dagger(x)\sqrt{\phi(x)},
        {d\over{dx}}\pv
        dy\,{{\psi(y)\sqrt{\phi(y)}}\over{x-y}}\biggr]  \cr
  }  \eqn \rescaledH
 $$
and for the supercharges
$$
  \eqalign{
    Q &\equiv \int dx\,\psi^\dagger(x)\sqrt{\phi(x)}\,
       \biggl(
       {{p(x)}} - {i\over {2\phi}}
       \left((\delx\psi^\dagger)\psi-\psi^\dagger (\delx \psi)\right)
           \cr
      &+ i\,\tilde v(x)
       - i \pv dy\,{{\dely\varphi}\over{x-y}}
        \biggr),
    \cr
    Q^\dagger &\equiv \int dx\,\psi(x)\sqrt{\phi(x)}\,
       \biggl(
       {{p(x)}} - {i\over {2\phi}}
       \left((\delx\psi^\dagger)\psi-\psi^\dagger (\delx \psi)\right)
           \cr
      &- i\,\tilde v(x)
       + i \pv dy\,{{\dely\varphi}\over{x-y}}
        \biggr),
    \cr
  } \eqn \rescaledQ
$$
with standard commutators
$$
  [\varphi(x),\varphi(y)] = 0, \quad [\varphi(x),p(y)] =
       i\,\delta(x-y),
  \quad \{\psi(x),\psi^\dagger(y)\} = \delta(x-y),
$$
$$
  [\varphi(x),\psi(y)] = [\varphi(x),\psi^\dagger(y)] =
  [\,p(x),\psi(y)] =
  [\,p(x),\psi^\dagger(y)] = 0.  \eqno \eq
$$
The square root factors appearing in equations \rescaledL\ to
\rescaledQ, when
expanded around the large $N$ background configuration, will generate
an infinite set of vertices.

This hamiltonian and these supercharges depend implicitly on $N$
through $v$ and the constraint $\int \phi = N$.  To make the
$N$-dependence explicit, observe that for $W$ of the form \generalW\
the expressions  $\tilde v (\sqrt N x)/\sqrt N$ and $\tilde v'(\sqrt N
x)$
are independent of $N$.  We are therefore motivated to rescale $x \to
\sqrt N x$.  Then to get rid of the $N$-dependence in the constraint,
we must rescale $\phi \to \sqrt N \phi$.

The complete rescaling is given by
$x \rightarrow\sqrt{N}\,x$, $\phi\rightarrow\sqrt{N}\,\phi$,
$\varphi\rightarrow N\varphi$, $p\rightarrow p/N^{3/2}$ and
$\psi^{(\dagger)} \rightarrow\psi^{(\dagger)}/N^{1/4}$.
Defining $v(x) \equiv \tilde v(\sqrt N x)/\sqrt N$, which is
{\it independent\/} of $N$, we get for the
supercharges
$$
  \eqalign{
    Q &\equiv \int dx\,\psi^\dagger(x)\sqrt{\phi(x)}\,
       \biggl(
       {{p(x)}\over N} - {i\over {2\phi N}}
       \left((\delx\psi^\dagger)\psi-\psi^\dagger (\delx \psi)\right)
           \cr
      &+ i\,N\,v(x)
       - i N \pv dy\,{{\dely\varphi}\over{x-y}}
        \biggr),
    \cr
    Q^\dagger &\equiv \int dx\,\psi(x)\sqrt{\phi(x)}\,
       \biggl(
       {{p(x)}\over N} - {i\over {2\phi N}}
       \left((\delx\psi^\dagger)\psi-\psi^\dagger (\delx \psi)\right)
           \cr
      &- i\,N\,v(x)
       + i N \pv dy\,{{\dely\varphi}\over{x-y}}
        \biggr),
    \cr
  } \eqn \rescaledQ
$$
and for the hamiltonian
$$
  \eqalign{
   H &=
       {{N^2} \over 2}\int dx\,\phi(x)
            \left(\pv dy\,{{\dely \varphi} \over
             {x-y}} - v(x) \right)^2 \cr
      &+ {1 \over {2N^2}} \int{{dx}\over\phi}\,
        \left(\phi p - {i\over 2} \left[(\delx\psi^\dagger)\psi
              -\psi^\dagger (\delx \psi)\right]\right)^2 \cr
      &- \half\int dx\, [\psi^\dagger, \psi] \,{d\over dx}\left(
              \pv dy\,{{\dely\varphi}\over{x-y}} - v(x)\right)  \cr
    &+ \half\int dx\,\biggl[\psi^\dagger(x)\sqrt{\phi(x)},
       {d\over{dx}}\pv
       dy\,{{\psi(y)\sqrt{\phi(y)}}\over{x-y}}\biggr].  \cr
  }  \eqn \rescaledH
 $$
The rescaled collective field satisfies the constraint
$$
  \int dx \,\phi (x) = 1.      \eqno \eq
$$

The leading term in the hamiltonian \rescaledH\ is just a constant.
If it is nonzero,
then supersymmetry is broken to leading order in $N$.  Conversely, if
the vacuum configuration of the density field satisfies
$$
  V_{\rm eff}(\phi_0) = {{N^2} \over 2}\int dx\,\phi_0(x)
     \left(\pv dy\,{{\phi_0(y)} \over {x-y}} - v(x)\right)^2 = 0.
  \eqno \eq
$$
then supersymmetry is preserved to leading order.
This will be the case if
$$
         \pv dy\,{{\phi_0(y)} \over
             {x-y}} - v(x) = 0.
  \eqn \nonFreeCond
$$
Throughout this paper we will assume that supersymmetry is unbroken at
the classical level and that equation \nonFreeCond\ applies.

To set up the perturbation theory,
we expand around the vacuum configuration $\phi_0$ as
$$
  \phi = \delx\varphi = \phi_0 + {1\over N}\,\delx\eta, \qquad
  p \to Np.  \eqno \eq
$$
The factor $1/N$ makes all bosonic
propagators of order unity in position space, and absorbs the explicit
$N$-dependence into the vertices.  This is purely a matter of
convenience, and the factor $1/N$ could be left out without changing
the theory, a fact that can be seen by power counting of graphs in
position space, or most simply by writing the respective
hamiltonians in terms of normalized creation and annihilation
operators and noting that the resulting expressions are identical.

Expanding about $\phi_0$, terms linear in $\partial\eta$ cancel
and the hamiltonian \rescaledH\ becomes
$$
 \eqalign{
   H &=
        \half \int dx\, \phi_0 p^2
         +{\pi^2\over 2} \int dx\,\phi_0(\delx\eta)^2
         -\half\int dx\pv dy\,{v(x)-v(y)\over
           x-y}\,\delx\eta\,\dely\eta   \cr
        &+ {1\over 2}\int
          dx\,\left[\psi^\dagger(x)\sqrt{\phi_0(x)},
          {d\over{dx}}\pv
          dy\,{{\psi(y)\sqrt{\phi_0(y)}}\over{x-y}}\right]\cr
     &+ {1\over{2N}}\int dx\,(\delx\eta)\,p^2
       + {1\over 2N}\int dx\, \delx\eta\left(\pv dy\,{\dely\eta\over
           x-y}\right)^2 \cr
     &- {i\over {2N}}\int dx\, \left[(\delx\psi^\dagger)\psi
           -\psi^\dagger (\delx \psi)\right] p   \cr
     &- {1\over{2N}}\int dx\,[\psi^\dagger(x),\psi(x)]\,
      {d\over dx} \pv dy\,{{\dely\eta}\over{x-y}}
         \cr
     &+ {1\over 2}\int
        dx\,\left[\psi^\dagger(x)\sqrt{\phi_0(x)}
        \left(\sqrt{{{\phi}/{\phi_0}}}(x)-1\right)
        ,
        {d\over dx}\pv
        dy\,{{\psi(y)\sqrt{\phi_0(y)}}\over{x-y}}\right]\cr
     &+ {1\over 2}\int
        dx\,\left[\psi^\dagger(x)\sqrt{\phi_0(x)},
        {d\over{dx}}\pv
        dy\,{{\psi(y)\sqrt{\phi_0(y)}}\over{x-y}}
        \left(\sqrt{{{\phi}/{\phi_0}}}(y)-1\right)
        \right]\cr
     &+ {1\over{8N^2}}\int{{dx}\over{\phi_0 + {1\over N}\delx\eta }}\,
        \left[\psi(\delx\psi^\dagger)
        (\delx\psi)\psi^\dagger + \psi^\dagger(\delx\psi)
        (\delx\psi^\dagger)\psi\right]  \cr
     &+ {1\over 2}\int
        dx\,\biggl[\psi^\dagger(x)\sqrt{\phi_0(x)}
        \left(\sqrt{{{\phi}/{\phi_0}}}(x)-1\right)
        ,    \cr
    &\qquad    {d\over dx}\pv
        dy\,{{\psi(y)\sqrt{\phi_0(y)}}\over{x-y}}
        \left(\sqrt{{{\phi}/{\phi_0}}}(y)-1\right)
        \biggr].\cr
 }   \eqn \expandH
$$

Some remarks are in order at this point:
In the bosonic collective field theory,
cubic terms such as appear in the bosonic part of the hamiltonian
\rescaledH\ are simplified using the identity
$$
 \int dx\,\phi(x)
   \left(\pv dy\,{\phi(y) \over {x-y}}\right)^2
 = {{\pi^2}\over 3}\int dx\,\phi^3(x),   \eqn \unregId
$$
which can readily be demonstrated [26] in Fourier space
using
$$
  \pv dy\, {{e^{iky}}\over{x - y}}
   = -\pi i \,\epsilon(k)\, e^{ikx}.  \eqno \eq
$$

In applying \unregId\ one must, however, be careful.
As it stands it cannot be valid if the integral
$\int \phi^3$ on the right hand side diverges.
This can indeed happen, both at the level of the vacuum density
$\phi_0$,
as happens in the potential free case (not discussed here) and at the
level of the fluctuations $\partial\eta$, as we shall see in the
harmonic
case.  To be general, we therefore {\it avoid\/} using
the identity
\unregId\ in this chapter, and treat issues of regularization later,
as they arise.

On the other hand, in \expandH\ we have rewritten the quadratic terms
by applying the  identity
$$
\eqalign{
 \int dx\,\phi_0(x)
   \left(\pv dy\,{\dely\eta \over {x-y}}\right)^2
   &+ 2\int dx\,\delx\eta
       \left(\pv dy\,{\phi_0(y) \over {x-y}}\right)
       \left(\pv dz\,{\delz\eta \over {x-z}}\right)  \cr
   &=
      {\pi^2} \int dx\,\phi_0(\delx\eta)^2,  \cr
} \eqno \eq
$$
which is easy to show in Fourier space.
For all the cases that we will consider, the integrals in \? are
well behaved and no special regularizations are needed.

Using
$$
  \sqrt{{{\phi}/{\phi_0}}}-1 = {1\over{2N\phi_0}}\,\delx\eta
    -{1\over{8N^2\phi_0^2}}\, (\delx\eta)^2 + o\left({1\over
    {N^2}}\right),     \eqno \eq
$$
the hamiltonian can be written up to cubic order as
$$
 \eqalign{
   H &=
      {1\over 2} \int dx\,\phi_0 p^2 + {{\pi^2}\over 2}\int dx\,
        \phi_0(\delx\eta)^2
         -\half\int dx\pv dy\,{v(x)-v(y)\over
           x-y}\,\delx\eta\,\dely\eta   \cr
    &+ {1\over 2}\int
          dx\,\left[\psi^\dagger(x)\sqrt{\phi_0(x)},
          {d\over{dx}}\pv
          dy\,{{\psi(y)\sqrt{\phi_0(y)}}\over{x-y}}\right]  \cr
     &+ {1\over{2N}}\int dx\,(\delx\eta)\,p^2
        + {{\pi^2}\over{6N}}\int dx\,(\delx\eta)^3
        - {i\over {2N}}\int dx\, \left[(\delx\psi^\dagger)\psi
           -\psi^\dagger (\delx \psi)\right] p \cr
     &- {1\over{2N}}\int dx\,[\psi^\dagger(x),\psi(x)]\,
         {d\over{dx}}\pv dy\,{{\dely\eta}\over{x-y}}
           \cr
     &+ {1\over{4N}}\int
        dx\,\left[{\psi^\dagger(x)\over\sqrt{\phi_0(x)}}\,
        \delx\eta
        ,
        {d\over{dx}}\pv
        dy\,{{\psi(y)\sqrt{\phi_0(y)}}\over{x-y}}\right]  \cr
     &+ {1\over{4N}}\int
        dx\,\left[{\psi^\dagger(x)\sqrt{\phi_0(x)}},
        {d\over{dx}}\pv
        dy\,{{\psi(y)/\sqrt{\phi_0(y)}}\over{x-y}}\,
        \dely\eta
        \right]
+ o\left({1\over N^2}\right),
\cr
 }   \eqn \perturbH
$$
and the supercharges  \rescaledQ\ can be written as
$$
 \eqalign{
   Q &=
        \int dx\,\sqrt{\phi_0}\, \psi^\dagger(x)
        \biggl\{
        p(x) - i\pv dy\,{{\dely\eta}\over{x-y}}
          \biggr\} \cr
     &\quad\, + {1\over{2N}}\int
        {dx\over\sqrt{\phi_0}}\,\psi^\dagger(x)
        \biggl\{\delx\eta\biggl[
        p(x) - i\pv dy\,{{\dely\eta}\over{x-y}}
        \biggr] \cr
     &\quad\, - i
        \left((\delx\psi^\dagger)\psi - \psi^\dagger(\delx\psi)\right)
        \biggr\} + o\left({1\over N^2}\right),             \cr
  Q^\dagger &= {\rm h.c.}. \cr
 }  \eqn \perturbQ
$$

We see that the system develops an infinite sequence of polynomial
interactions in the bare string coupling constant $1/N^2$.  This is in
contrast to the cubic bosonic collective field theory hamiltonian [13,
20] and
does not depend on the presence of a potential $v$.  The supercharges
also acquire expansions to all orders in perturbation theory, typical
of supersymmetric theories expanded about background configurations.
In general, the presence of a nontrivial background
$\phi_0$ may  \lq\lq
dress\rq\rq\ the coupling constant and possibly allow us to take a
suitable double scaling limit.

One of the remarkable properties of the
cubic bosonic hamiltonian is that [20] the integral representation
of a given amplitude can in general be reinterpreted as a sum of
standard tachyon exchange diagrams, plus contact terms, in a one to
one correspondence to first quantized Liouville computations [25].
In studies of critical closed string field theory, an infinite
sequence of polynomial interactions seems to be required to obtain
agreement with the first quantized integrations over moduli space
[34].  In the model discussed in this paper, the need to also
include an
infinite set of vertices is unavoidable.  It should
be clear that the reason for the supersymmetrized version of the
 bosonic cubic hamiltonian to contain an infinite sequence of
higher order vertices with derivative couplings is that the
supercharges themselves have an infinite expansion.

It is conceivable that other supersymmetric extensions of the
bosonic collective
string field hamiltonian (possibly formulated directly in the
continuum) may exist, with properties different from those described
here.  Ultimately, once a genuine field theory of $d=1$ superstrings
is formulated, the correct choice would be selected by requiring
agreement with the super-Liouville theory [6].

\chapter {SPECTRUM}

\section {Semiclassical spectrum}

\noindent
In the super-Calogero description of the model [15] it is
straightforward to show that if a supersymmetric classical
configuration can be found, the semiclassical spectrum is
supersymmetric.  A similar argument is developed here in the continuum
description of the model.

Referring back to the previous chapter, we note that
the condition \nonFreeCond\ for supersymmetry at the level of the
classical vacuum  can equivalently be restated as
$$
  \left. W_{;x} \right|_{\phi_0} = 0,  \eqno\eq
$$
where $W$ stands for the effective superpotential of
\generalContW.
When this condition is satisfied,
the continuum hamiltonian \continuumH\ can
be expanded around $\phi_0$ to give a quadratic contribution
$$
 H_0 = H_0^B + H_0^F,
\eqno\eq
$$
where
$$
\eqalign{
  H_0{}^B &= \half \int dx\,\phi_0 (x)\left(p^2
   + \int dy\int dz\, W_{;xy}W_{;xz}\,\eta(y)\,\eta(z)\right), \cr
  H_0{}^F &= \half\int dx\int dy\,
    \left[\psi^\dagger(x), \sqrt{\phi_0(x)}\, W_{;xy}\,
          \sqrt{\phi_0(y)} \psi(y)
  \right],    \cr
}   \eqno \eq
$$
and where the expressions of the form $W_{;xy}$ are evaluated at
$\phi_0$.

Defining the change of variables
$dq = dx/\phi_0$ and rescaling $p\to p/\phi_0$, $\psi \to
\psi/\sqrt{\phi_0}$, $\psi^\dagger\to\psi^\dagger/\sqrt{\phi_0}$, the
hamiltonian simplifies to give
$$
\eqalign{
  H_0{}^B &= \half \int dq\,\left(p^2
   + \left(\int dq'\; W_{;qq'}\,\eta(q')\right)^2\right), \cr
  H_0{}^F &= \half\int dq\,
    \left[\psi^\dagger(q), \int dq'\, W_{;qq'}\,
           \psi(q')
  \right],            \cr
}   \eqno \eq
$$
where we have used the identity $\delta/\delta\varphi(q) =
\phi_0(q)\,\delta/\delta\varphi(x)$, which follows by the chain rule,
keeping in mind the fact that $\delta (q-q') = \phi_0
(q)\,\delta(x-x')$.

Now let $\tilde W$ be the kernel defined by
$$
\tilde W (\phi)(q) \equiv \int dq'\,W_{;qq'\,}\phi(q').  \eqno\eq
$$
Then the quadratic hamiltonian can be written as
$$
\eqalign{
  H_0{}^B &= \half \int dq\,\left(p^2
   + \left(\tilde W(\eta)(q)\right)^2\right), \cr
  H_0{}^F &= \half\int dq\,
    \left[\psi^\dagger(q), \tilde W
           (\psi)(q)
  \right],            \cr
}   \eqn \kernelH
$$

Let $\{\phi_n\}$ be a complete set of normalised eigenfunctions of
$\tilde W$, \ie,
$$
  \tilde W \phi_n = \lambda_n\phi_n.  \eqno \eq
$$
Assuming $\tilde W$ to be positive definite, we can expand the fields
as
$$
\eqalign{
  \eta &= \sum_n\, {1\over\sqrt{2\lambda_n}}\,
           (a_n \phi_n + a^\dagger_n \phi^*_n),   \cr
  p    &= \sum_n\, -i\sqrt{\lambda_n\over 2}\,
           (a_n \phi_n - a^\dagger_n \phi^*_n),   \cr
  \psi &= \sum_n\, b_n\phi_n,   \cr
  \psi^\dagger &= \sum_n\, b^\dagger_n\phi^*_n,   \cr
}  \eqno \eq
$$
where $[a_m, a^\dagger_n] = \delta_{mn}$ and $\{b_m,
b^\dagger_n\}=\delta_{mn}$, all other commutators vanishing.

It is then trivial to show that
$$
\eqalign{
  H_0{}^B &= \sum_n\, \lambda_n (a_n^\dagger a_n + \half),  \cr
  H_0{}^F &= \sum_n\, \lambda_n (b_n^\dagger b_n - \half),  \cr
}  \eqno \eq
$$
thus demonstrating explicit supersymmetry of the quadratic spectrum.

Restricting attention to the bosonic piece, in addition to the terms
associated in the bosonic string theory [13] to a massless
scalar particle, one has the additional contribution
$$
         -\half\int dx\pv dy\,{v(x)-v(y)\over
           x-y}\,\delx\eta\,\dely\eta,
$$
which may in general affect
the dynamics in unexpected ways.  However, using the fact that $\int
dx\, \delx\eta = 0$, it is easy to show that for potentials of the
general form $v(x) = a + bx + cx^2$, this term falls away and the
quadratic spectrum is that of a massless scalar. These potentials
include the free case, the harmonic
case, and those of references [10, 24].  We will also be interested in
a potential of the general form $v(x) = bx + d x^3$.  In this case a
symmetric ansatz [14] can be consistently implemented and again the
bosonic quadratic spectrum is that of a massless scalar.

Explicitly, in these cases the quadratic part of the bosonic sector
of the hamiltonian \perturbH\ reads
$$
  H_0{}^B = {1\over 2} \int dx\,\phi_0 p^2 + {{\pi^2}\over 2}\int dx\,
        \phi_0(\delx\eta)^2,
\eqno\eq
$$
or, in $q$-space,
$$
  H_0{}^B = {1\over 2} \int dq\,\left( p^2 + {{\pi^2}}
        (\delq\eta)^2\right).   \eqno\eq
$$

In the light of the above discussion, we already know that assuming
that there exists a supersymmetric classical configuration, the
fermionic spectrum will also be that of a massless scalar particle.
One should therefore be able to rewrite the fermionic piece
$$
\eqalign{
  H_0{}^F &= {1\over 2}\int
          dx\,\left[\psi^\dagger(x)\sqrt{\phi_0(x)},
          {d\over{dx}}\pv
          dy\,{{\psi(y)\sqrt{\phi_0(y)}}\over{x-y}}\right]  \cr
&= \half\int
   dq\,\left[\psi^\dagger(q),
   {d\over{dq}}\pv
   dq'\,{{\phi_0(q')\psi(q')}\over{x(q)-x(q')}}\right]   \cr
}\eqno\eq
$$
in a form in which this property is manifest.

Assuming $q$ to be defined on $[0, L]$, we expand
$\eta$ and $p$ as
$$
\eqalign{
  \eta(q) &= \sum_{n=1}^\infty\, {1\over\sqrt{2\pi^2 n}}\,
          (a_n + a_n^\dagger)\, \sin{n\pi q\over L},  \cr
  p(q)  &= \sum_{n=1}^\infty\, -i\sqrt{\pi^2 n\over 2L^2}\,
          (a_n - a_n^\dagger)\, \sin{n\pi q\over L},  \cr
}   \eqn\sineExpansion
$$
where $[a_m,a_n^\dagger]=\delta_{mn}$.
In the above Dirichlet boundary conditions have been assumed.

The bosonic part of the quadratic hamiltonian is then simply given by
$$
  H_0{}^B = \sum_{n=0}^\infty\,{n\pi^2\over L}\,(a_n^\dagger a_n +
\half).
  \eqn\sineH
$$

In \kernelH\ the hamiltonian was expressed in terms of a kernel
$\tilde W$ as
$$
\eqalign{
  H_0{}^B &= \half \int dq\,\left(p^2
   + \left(\tilde W(\eta)(q)\right)^2\right), \cr
  H_0{}^F &= \half\int dq\,
    \left[\psi^\dagger(q), \tilde W
           (\psi)(q)
  \right].            \cr
}   \eqno \eq
$$
The same kernel therefore appears in both the bosonic and the
fermionic part of the hamiltonian.
Remembering that we want to rewrite the fermionic piece in a simpler
form, our motivation is therefore to solve for $\tilde W$ by
determining its properties.

Expanding with respect to a complete set of normalised
Dirichlet eigenfunctions of $\tilde W$ on $[0,L]$, we had
$$
\eqalign{
  \eta &= \sum_n\, {1\over\sqrt{2\lambda_n}}\,
           (\tilde a_n \phi_n + \tilde a^\dagger_n \phi^*_n),   \cr
  p    &= \sum_n\, -i\sqrt{\lambda_n\over 2}\,
           (\tilde a_n \phi_n - \tilde a^\dagger_n \phi^*_n),   \cr
}  \eqn \tildeExpansion
$$
and
$$
\eqalign{
  H_0{}^B &= \sum_n\, \lambda_n (\tilde a_n^\dagger \tilde a_n +
\half). \cr
}  \eqn\tildeH
$$

Comparing \tildeH\ and \sineH, one immediately sees that in the
present case $\tilde W$ has eigenvalues $\lambda_n = \pi^2 n/L$.
All that remains is therefore to explicitly solve for its
eigenfunctions $\phi_n$.

As both \sineExpansion\ and \tildeExpansion\ are expansions of
the same fields in terms of normalised oscillator coordinates, it
follows that the $a$'s are unitarily related to the $\tilde a$'s,
\ie,
$$
  a_n = U_{ni} \tilde a_i,  \qquad a^\dagger_n = U^*_{ni}
    \tilde a^\dagger_i,   \eqno\eq
$$
where $U^\dagger U=1$.  Comparing \tildeH\ and
\sineH\ one finds
$$
 \sum_n\, \lambda_n \tilde a_n^\dagger \tilde a_n
 = \sum_{n\,i\,j}\, \lambda_n\, U^*_{nj}\,\tilde a_j^\dagger\,
     U_{ni}\,\tilde a_i,
  \eqno\eq
$$
from which it follows by comparing coefficients that
$U^\dagger\Lambda U = \Lambda$, where $\Lambda$ is the diagonal matrix
with $\Lambda_{ii} = \lambda_i$.  Using the fact that $U$ is unitary,
this is equivalent to $[U, \Lambda] = [U^\dagger, \Lambda]= 0$.  Thus
$U$ leaves the eigenspaces of $\Lambda$ invariant, and as the
eigenvalues $\lambda_n = \pi^2 n/L$ are nondegenerate, we conclude
that $U$ is diagonal with complex phase factors on the diagonal.

Thus each $a_n$ is related to $\tilde a_n$ by a
phase factor.  Comparing \sineExpansion\ and \tildeExpansion, it
follows that the eigenfunctions of $\tilde W$ are given by
$$
  \phi_n(q) = {1\over\sqrt L} \sin {\pi n q\over L},
\eqno\eq
$$
modulo an inessential phase. In other words,
$$
  \tilde W \sin {\pi n q\over L} = {\pi^2 n\over L}\sin {\pi n q\over
   L}. \eqno\eq
$$

Using the identity
$$
 \pv dq\, {e^{ikq}\over q'-q}= -\pi i\,\epsilon(k)\,e^{ikq'},
\eqno\eq
$$
one  verifies that
$$
 {d\over dq'}\pv dq\, {\sin{kq}\over q'-q}= \pi \,|k|\,
       \sin{kq'}, \eqno\eq
$$
so that the kernel can be written as
$$
  \tilde W (\phi) (q') =
 {d\over dq'}\pv dq\, {\phi{(q)}\over q'-q}
        \eqno\eq
$$

The conclusion is therefore that the fermionic piece
of the hamiltonian \kernelH\ can be written as
$$
\eqalign{
  H_0{}^F &= \half\int dq\,
    \left[\psi^\dagger(q), \tilde W
           (\psi)(q)
  \right],            \cr
   &= \half \int dq\,
    \left[\psi^\dagger(q),
           {d\over dq}\pv dq'\, {\psi{(q')}\over q-q'}
  \right].            \cr
}   \eqno \eq
$$
In other words, for the class of potentials $v(x)$ described
above,
the quadratic
hamiltonian can be rewritten in a form manifestly that of a free
Majorana fermion.  We have also proved the identity
$$
\int
   dq\,\psi^\dagger(q)\,
   {d\over{dq}}\pv
   dq'\,{{\phi_0(q')\psi(q')}\over{x(q)-x(q')}}
= \int dq\,
  \psi^\dagger(q) \,
       {d\over dq}\pv dq'\, {\psi{(q')}\over q-q'},
\eqno\eq
$$
for the same class of potentials.  This
identity may look deceptively trivial to prove via contour integration
arguments, due to the fact that $\phi_0 = dx/dq$, so that both
integrands
have the same residues where both have poles.  However, in general the
function $x$ may have
additional poles in unfavourable positions for the naive argument to
be valid.

\chapter {THE HARMONIC CASE}

\noindent
In this chapter, we study the model characterised by
the effective superpotential
$$
  W = \int dx\,\bar W(x)\,\delx\varphi - {1\over 2}
     \int dx \int dy\, \ln |x-y|\,\delx\varphi\,\dely\varphi.
     \eqno \eq
$$
in the case where $\bar W(x)={\omega\over 2}x^2$.  As remarked
before (see \generalContW), this model is a continuum version of the
original super-Calogero system \CalogeroW\ and \sCalogeroH.
Equivalently, it can be seen as a continuum formulation of a matrix
model of the form \matrixS, where the matrix superpotential
$\bar W(\Phi)={\omega\over 2}\Phi^2$ is purely
harmonic.  This system has been studied in the super collective field
formalism before in [14], and an expanded version of the
analysis will be presented here.

\section {Vacuum density and spectrum}

\noindent
For supersymmetry to be
unbroken to leading order in $N$
the vacuum configuration of the density field must satisfy
$$
         \pv dy\,{{\phi_0(y)} \over
             {x-y}} - \omega x = 0,
  \eqn \vacCondition
$$
where $\phi_0$ must satify the constraint $\int\phi_0 = 1$.

We can use the identity \unregId\ to
express the effective bosonic potential
$$
  V_{\rm eff}(\phi) = {{N^2} \over 2}\int dx\,\phi(x)
     \left(\pv dy\,{{\phi(y)} \over {x-y}} -
     \omega x\right)^2
  \eqno \eq
$$
of the hamiltonian \rescaledH\ in the form
$$
  V_{\rm eff}(\phi)
   = N^2 \left( {{\pi^2}\over 6}\int \phi^3
     -{\omega\over 2}\left(\int\phi\right)^2 + {\omega^2\over 2} \int
x^2\phi
         \right).   \eqno\eq
$$
The background $\phi_0$ satisfies the stationarity condition
$$
  {{\pi^2\phi_0^2}\over 2} - \omega \int\phi_0 + {\omega^2\over 2}x^2
   = 0.
  \eqn\harmStationarity
$$
Thus one gets
$$
  \phi_0 (x) = {1\over\pi}\sqrt {\mu_F - \omega^2 x^2},
  \eqno \eq
$$
where $\mu_F \equiv  2\omega\int\phi_0$.
Requiring that $\int\phi_0 = 1$, one finds $\mu_F = 2\omega$,
which is consistent with the standard analytic solutions of
\vacCondition.

We recall that the quadratic piece of the hamiltonian \perturbH\ is
$$
\eqalign{
  H_0 &=
       {1\over 2} \int dx\,\phi_0 p^2 + {{\pi^2}\over 2}\int dx\,
        \phi_0(\delx\eta)^2   \cr
    &+ {1\over 2}\int
          dx\,\left[\psi^\dagger(x)\sqrt{\phi_0(x)},
          {d\over{dx}}\pv
          dy\,{{\psi(y)\sqrt{\phi_0(y)}}\over{x-y}}\right]   \cr
}  \eqno\eq
$$
or, after
changing variables
$dq = dx/\phi_0$ and rescaling $p\to p/\phi_0$, $\psi \to
\psi/\sqrt{\phi_0}$, $\psi^\dagger\to\psi^\dagger/\sqrt{\phi_0}$,
$$
\eqalign{
  H_0 &=
       {1\over 2} \int dq\, p^2 + {{\pi^2}\over 2}\int dq\,
        (\delq\eta)^2   \cr
    &+ {1\over 2}\int
          dq\,\left[\psi^\dagger(q),
          {d\over{dq}}\pv
          dq'\,{{\phi_0(q')\psi(q')}\over{x(q)-x(q')}}\right]   \cr
    &=
       {1\over 2} \int dq\, p^2 + {{\pi^2}\over 2}\int dq\,
        (\delq\eta)^2   \cr
    &+ {1\over 2}\int
          dq\,\left[\psi^\dagger(q),\phi_0(q)
          \pv
          dq'\,{{\partial_{q'}\psi}\over{x(q)-x(q')}}\right].
\cr }  \eqn\harmHzero
$$

Explicitly, the above change of variables reads
$$
\eqalign{
  q &= \pi \int {dx\over\sqrt{\mu_F-\omega^2 x^2}}  \cr
    &= {\pi\over\omega}\arccos {-x\over a},
}   \eqno\eq
$$
where $q$ has been chosen to be zero at the turning point
$x = -a$,
$a = \sqrt{\mu_F}/\omega = \sqrt {2/\omega}$.
Inverting \?, one finds
$$
  x =
      -\sqrt{2L\over\pi^2} \cos{\pi q\over L}; \qquad q\in [0,L],
$$
where $L = {\pi^2/\omega} = \pi^2 a^2/2$ is the half period.

Expanding $\eta$, $p$ and $\psi^{(\dagger)}$ on $[0,L]$ as
$$
\eqalign{
  \eta(q) &= \sum_{n=1}^\infty\, {1\over\sqrt{\pi^2 n}}\,
          (a_n + a_n^\dagger)\, \sin{n\pi q\over L},  \cr
  p(q)  &= \sum_{n=1}^\infty\, -i\sqrt{\pi^2 n\over L^2}\,
          (a_n - a_n^\dagger)\, \sin{n\pi q\over L},  \cr
   \psi^{(\dagger)}(q) &= {\sqrt {2\over L}}\,\sum_{n=1}^\infty\,
   b_n^{(\dagger)}\,
         \sin {n\pi q\over L},   \cr
}  \eqno \eq
$$
where $[a_m,a_n^\dagger]=\delta_{mn}$ and
$\{b_m, b^\dagger_n\}=\delta_{mn}$,
the bosonic part of the quadratic hamiltonian becomes
$$
  H_0{}^B = \sum_{n=1}^\infty\,n\omega\,(a_n^\dagger a_n + \half).
  \eqno\eq
$$

For the fermionic part, one needs to evaluate
$$
\eqalign{
     \pv_{\!\!\!\! 0}^L
          dq'\,{{
            \cos{\pi n q'\over L}}\over{\cos{\pi q'\over
                      L}-\cos{\pi q\over L}}}
     &= \half\,
     \pv_{\!\!\!\! -L}^L
          dq'\,{{
            \exp{i\pi n q'\over L}}\over{\cos{\pi q'\over
                      L}-\cos{\pi q\over L}}},
}  \eqno \eq
$$
This can be done by closing the contour in the upper half plane,
exploiting the fact that the integrand is periodic with period $2L$.
The only relevant poles are on the real line at $+q$ and $-q$.  The
principal
value prescription corresponds to taking half of the residues of any
poles that lie on the contour of integration.  Applying this, one
obtains
$$
     \pv_{\!\!\!\! 0}^L
          dq'\,{{
            \cos{\pi n q'\over L}}\over{\cos{\pi q'\over
                      L}-\cos{\pi q\over L}}}
   =  L\, \epsilon (n)\,{\sin {\pi n q\over L}\over\sin {\pi  q\over
L}}.
\eqn\contourId
$$
Using this result, $H_0{}^F$ is now easily rewritten in the form
$$
  H_0{}^F = \sum_{n=1}^\infty \,n\omega\, (b^\dagger_n b_n-\half),
   \eqno \eq
$$
thus explicitly demonstrating supersymmetry of the semiclassical
spectrum.

We notice that equation \contourId\ is all that is required for both
the spectrum and the ground state configuration.  Indeed, it is
straightforward to verify that equation \vacCondition\ is
self-consistently satisfied provided one defines $\epsilon (n=0) = 0$.
Equation \contourId\ is also related to an identity involving
Chebychef polynomials of the second type [14].

\section {Three point functions}

\noindent
We now compute the supercharges and three point functions of the
harmonic theory in the oscillator basis.

Using \vacCondition\ and changing variables to $q$-space as in the
previous section, the supercharges \perturbQ\ become
$$
 \eqalign{
   Q
     &=
        \int_0^L dq\, \psi^\dagger(q)
        \biggl\{
        p(q) - i\,\phi_0(q)\pv_{\!\!\!\! 0}^L
              dq'\,{{\delqp\eta}\over{x(q)-x(q')}} \biggr\} \cr
     &\quad + {1\over{2N}}\int_0^L
        {dq\over{\phi_0^2}}\,\psi^\dagger(q)\,
        \delq\eta\left\{
          p(q) - i\,\phi_0(q)
            \pv_{\!\!\!\! 0}^L
              dq'\,{{\delqp\eta}\over{x(q)-x(q')}}\right\} \cr
     &\quad - {i\over 2N}\int_0^L
        {dq\over{\phi_0^2}}\,\psi^\dagger
        (\delq\psi^\dagger)\psi
        + o\left({1\over N^2}\right),                 \cr
  Q^\dagger &= {\rm h.c.}. \cr
 }  \eqn \qspaceQ
$$

For the quadratic term in $Q$ one uses equation \contourId\ to readily
obtain
$$
  Q_0 = -i\sqrt{2\pi}\sum_{m>0}\, \sqrt{k_m}\, b^\dagger_m
  a_m.    \eqn \Qzero
$$

For the cubic terms in $Q$ one needs to compute the integrals
$$
  \pv_{\!\!\!\! -L}^L dq\, {e^{ik_n q}\over\phi_0^2(q)}.
$$
For $n>0$ this can be done by extending the contour in the upper half
plane vertically upwards at $-L$ and $L$. Exploiting the fact that the
integrand is periodic with period
$2L$, it follows that the vertical contributions cancel.  For $n<0$
one can similarly close the contour in the lower half plane. The
function
$1/\phi_0^2 = L / 2 \sin^2(\pi q/L)$ has singularities on the
contour at $-L$, at $0$ and at $L$.
Applying a principal part
regularization, which corresponds to taking half of the residues of
any poles
that lie fully on the contour and $1/4$ of the residues of poles that
are located at the corners $-L$ and $L$, one finds that for $n$ odd
the integral vanishes while for $n$ even it is given by
$$
  \pv_{\!\!\!\! -L}^L dq\, {e^{ik_n q}\over\phi_0^2(q)} =
    - {L^3\over\pi}\, |k_n|.     \eqn\kernelB
$$

Using this result one finds, after some algebra, that the
cubic piece of $Q$ is given by
$$
\eqalign{
  Q_3 &= {iL\sqrt L\over 8\pi N}\,
    {\sum_{m,n,p\ne 0}}{}^{\!\!\!\!\! \prime}\,
\epsilon(n)\sqrt{|k_nk_p|}\,|k_m+k_n+k_p|
      \,(b^\dagger_{-m}a_n a_p- b^\dagger_{-m}a^\dagger_{-n}a_p) \cr
  &+ {iL\sqrt L\over 8\pi N}\,
    {\sum_{m,n,p\ne 0}}{}^{\!\!\!\!\! \prime}\,k_n|k_m+k_n-k_p|\,
      b^\dagger_m b^\dagger_n b_p,    \cr
}  \eqn\Qthree
$$
where the primes on the sums indicate that one only sums over
indices such that the arguments of the absolute value signs are even.

The cubic piece of the hamiltonian can now be generated from \Qzero\
and \Qthree\ using $H = \half\{Q,Q^\dagger\}$.  One finds, after some
algebra, that
$$
  H_3 = H_3{}^B + H_3{}^F,
$$
where
$$
\eqalign{
  H_3{}^B &= {L\over 4N}\sqrt{L\over 2\pi}\,
            {\sum_{n,p,l>0}}{}^{\!\!\!\!\! \prime}\,\sqrt{k_l k_n k_p}\,
            \bigl(|k_l+k_n+k_p|    \cr
          &-|k_l-k_n-k_p|\bigr)\,
            a^\dagger_n a^\dagger_p a_l + {\rm h.c.},   \cr
  H_3{}^F &= {L\over 8N}\sqrt{L\over 2\pi}\,
             {\sum_{n,p,l>0}}{}^{\!\!\!\!\! \prime}\,\sqrt{k_l}\,(k_p-k_n)\,
             \bigl(|k_l+k_n+k_p|-|k_l+k_n-k_p|  \cr
          &\quad\,  +|k_l-k_n+k_p|-|k_l-k_n-k_p|\bigr)\,
             b_n^\dagger b_p a_l^\dagger + {\rm h.c.}   \cr
}   \eqn \Hthree
$$

\chapter {STATIONARY POINT AND D=1 STRINGS}

\noindent
We discuss here whether it is possible to identify in this model a
continuum limit which, in the bosonic sector of the theory, reduces to
the standard continuum limit of $d=1$ bosonic strings [2].

Consider the bosonic part of the potential in \rescaledH:
$$
  V_{\rm eff}(\phi) = {{N^2} \over 2}\int dx\,\phi(x)
     \left(\pv dy\,{{\phi(y)} \over {x-y}} - v(x)\right)^2,
  \eqno \eq
$$
where $v(x)$ is an arbitrary polynomial.
Since $\phi$ is positive, a necessary and
sufficient condition for supersymmetry to be preserved to leading
order in $N$ is that there exist a classical configuration $\phi_0$
such that
$$
         \pv dy\,{{\phi_0(y)} \over
             {x-y}} - v(x) = 0,
  \eqn \condA
$$
with $\int\phi_0 = 1$.
There is a well-known way of solving this equation using analytic
methods [1].  These solutions correspond to $d=0$ matrix models [10,
36].

What characterizes the standard formulation of the continuum limit of
bosonic $d=1$ strings is that the theory can be scaled in such a way
that
(a) the potential depends on a chemical potential $\mu$ (fermi
energy); (b) this chemical potential is a free parameter since all
dependence on the matrix coupling constant $g$ (which we will refer to
as the cosmological constant) is scaled out with the normalization
condition becoming $g=\int \phi(x)\,dx$.

As is well known, in the collective field theory approach
a lagrange multiplier (chemical potential) is introduced to enforce
the constraint $\int \phi(x)\,dx = 1$.  Therefore let us consider
$$
 \tilde V_{\rm eff}(\phi)/N^2 = V_{\rm eff}(\phi)/N^2
     - \mu \left(\int \phi - 1\right).
  \eqno \eq
$$
Notice that $\tilde V_{\rm eff}$ is not positive definite any longer.
If we require $\tilde V_{\rm eff}$ to be stationary with respect to
$\phi$, \ie,
$$
\delta_\phi \tilde V_{\rm eff}
    = \delta_\phi V_{\rm eff} - \mu = 0,   \eqn \refB
$$
it immediately follows that if supersymmetry is to be preserved (if
equation \condA\ is to be satisfied), then $\mu=0$.   This argument
would seem to immediately rule out the possibility of introducing an
arbitrary parameter $\mu$.

However, the argument is not that simple.  The correct equation of
stationarity is
$$
  \delta_\varphi \tilde V_{\rm eff}
    = \partial_x\,\delta_\phi \tilde V_{\rm eff} = 0.   \eqno \eq
$$
Indeed, this is also the case in the standard collective field theory
description of $d=0$ strings, and it is known that the above equation
is the statement of stationarity with respect to the original (\lq\lq
master\rq\rq) variables of the theory [21].  This equation is
compatible with the presence of a lagrange multiplier.  But, if
$\partial_x\,\delta_\phi \tilde V_{\rm eff} = 0$, this implies that
$\delta_\phi \tilde V_{\rm eff} = {\rm constant} \equiv c$.  So we
recover equation \refB\ with $\mu \to \mu-c$.  However, from our
discussion following equation \refB, for
supersymmetry to be preserved the equation of stationarity remains
$$
  \delta_\phi V_{\rm eff} = 0  \eqno \eq
$$
without allowing for an extra parameter to be added.

We now use the identity \unregId\ to rewrite this last equation as
$$
  {\pi^2\over 2}\, \phi_0^2(x)
      + \half\, v^2(x)
      - \pv dy \,{v(x)-v(y)\over x-y}\,\phi_0(y) = 0.
  \eqn \condB
$$
When the classical vacuum configuration is supersymmetric then
the last term in \condB\ may generate an effective chemical
potential, an example of which we saw in the harmonic case (see
equation \harmStationarity\ and the subsequent remarks).
This should not be too surprising, since the universal genus zero
contribution to the ground state energy of the $d=1$ bosonic string
theory [2] is different from zero and therefore one should presumably
not expect $g\to g_c$ with $g_c$ independent of $\triangle \mu$.

Let us now consider the scaling behaviour of equation \condB.
We recall that the bosonic
analogue of the stationarity
condition \condB\ is given by
$$
  {\pi^2\over 2}\, \phi_0^2(x)
      +  w(x)  - \mu = 0,
  \eqn \condC
$$
with the constraint $\int \phi_0 = 1$.
The bosonic potential $w$ is in general chosen to depend on a
coupling constant $g$ in such a way that $w(x/\sqrt g) =
\tilde w(x)/g$, where the rescaled potential $\tilde w$ has no
explicit
dependence on $g$.  Rescaling $x\to x/\sqrt g$, $\phi\to \phi/\sqrt
g$ and $\mu \to \mu/g$, all explicit dependence on $g$ scales
out of the stationarity condition, which becomes
$$
  {\pi^2\over 2}\, \phi_0^2(x)
      + \tilde w(x)  - \mu = 0,
  \eqn \condD
$$
while the constraint becomes $\int \phi_0 = g$.  We see that by
changing the coupling constant $g$, which now appears in the
normalization condition, we can freely adjust
the fermi level $\mu$ with respect to the rescaled potential
$\tilde w$.  In particular, as $\mu$ approaches a $-x^2$ maximum, the
time of flight $L \propto \int dx/\phi_0$ becomes infinite.

Let us now try to duplicate this in the fermionic case.
Taking
the potential in \condB\ to have the general scaling $v(g^a x) = g^b
\tilde v(x)$, where $\tilde v$ is independent of $g$ (note that this
is always the case for a monomial), and
rescaling $x\to g^ax$, $\phi\to g^c\phi$ and $\mu\to g^{2c}\mu$,
the stationarity condition \condB\ becomes
$$
  g^{2c}\,{\pi^2\over 2}\, \phi_0^2(x)
      + g^{2b}\, \half\, \tilde v^2(x)  - g^{2c}\mu
      - g^{b+c}\pv dy \,{\tilde v(x)-\tilde v(y)\over
x-y}\,\phi_0(y) = 0,
  \eqno\eq
$$
while the constraint becomes $\int \phi_0 = g^{-a-c}$.
For generality we have reintroduced a finite $\mu$, allowing for
supersymmetry breaking vacuum configurations.
Expanding $\tilde v$ about
$x$ and using this constraint, the stationarity condition becomes
$$
\eqalign{
  &
  0 = g^{2c}\,{\pi^2\over 2}\, \phi_0^2(x)
      + g^{2b}\, \half\, \tilde v^2(x)  - g^{2c}\mu  \cr
  &\quad
      - g^{b+c-a-c} \,\tilde v' (x)
      + \half\,g^{b+c-a-c}\, \tilde v'' (x)\,x
      - \half\,g^{b+c}\, \tilde v'' (x) \int dy\,y\, \phi_0(y)  \cr
  &\quad   + \dots.  \cr
}  \eqno\eq
$$
We see that if we choose $a = -b =-c$, an overall factor $g^{b+c}$
divides out of the equation and we find
$$
\eqalign{
  &
  0 = {\pi^2\over 2}\, \phi_0^2(x)
      +  \half\, \tilde v^2(x)  - \mu  \cr
  &\quad
      - \tilde v' (x)
      + \half\ \tilde v'' (x)\,x
      - \half\, \tilde v'' (x) \int dy\,y\, \phi_0(y)  \cr
  &\quad   + \dots,  \cr
}  \eqno\eq
$$
while the constraint is given by $\int\phi_0 = g^{-a-c} = 1$.  We have
thus succeeded in getting rid of all explicit and implicit dependence
of $\phi_0$ on $g$.  Rewriting this we therefore get
$$
  {\pi^2\over 2}\, \phi_0^2(x)
      + \half\, \tilde v^2(x)  - \mu
      - \pv dy \,{\tilde v(x)- \tilde v(y)\over x-y}\,\phi_0(y) = 0,
  \eqno \eq
$$
with the constraint $\int\phi_0 = 1$.  A point we want to
emphasize is that in contrast to the bosonic case the parameter $g$
does not appear
in the constraint, so that the
effective fermi level is not a
free parameter relative to the rescaled effective potential in \? and
cannot be adjusted to approach a maximum.  (The effective fermi level
is given by $\mu$ plus any $x$-independent terms arising from the
cross term in \?). The time of flight,
given by $L \propto \int dx/\phi_0$, remains finite and no true
extra dimension is generated.

The above argument assumes that $v(x)$ is homogeneous of arbitrary
order in only one parameter $g$,
and does not make a statement on the
possibility or not of reproducing an analogue of the bosonic scaling
limit when this is not the case.  This seems very hard to do in
specific examples that have been considered (not reproduced here),
but more work needs to be done to gain a better understanding of this
issue.

\chapter {TURNING POINT DIVERGENCES}

\noindent
We now discuss the regularization of turning point contributions as
it applies to the calculation of the three point functions of
chapter 5.

In chapter 3 it was mentioned that there may in general be subtleties
in the application of the identity \unregId
$$
 \int dx\,\delx\eta
   \left(\pv dy\,{\dely\eta \over {x-y}}\right)^2
 = {{\pi^2}\over 3}\int dx\,(\delx\eta)^3
\eqn\xspaceId
$$
given in $q$-space by
$$
   \int {dq}\,\delq\eta
            \left(\pv_{\!\!\!\! 0}^L
              dq'\,{{\delqp\eta}\over{x(q)-x(q')}}\right)^2   =
   {\pi^2\over 3}\int {dq\over\phi_0^2}\,(\delq\eta)^3,
\eqn\qspaceId
$$
used to simplify the
cubic piece of the hamiltonian.

In the calculation of the oscillator expansions
in chapter 5, we effectively started from
the left hand side of \qspaceId.  Our regularization scheme
consisted of taking principal parts at the poles.
On the other hand, the authors of [20] started from
the right hand side of \qspaceId, their regularization also coming
down to taking principal
parts. Comparing the bosonic part of our hamiltonian \Hthree\
to the one that
was obtained in [20] for the bosonic
collective field theory, we see that our expression differs from
theirs in that we have no terms of the form $a_m a_n a_p$.

To see this difference more clearly, note that
starting from the left hand side of \qspaceId,
assuming that $\eta$ obeys Dirichlet boundary conditions so that we
can expand $\delq\eta = \sum_{n>0}\,\phi_n\cos k_n q$, and using
the principal part prescription, one finds
$$
\eqalign{
   \int {dq}\,\delq\eta
            &\left(\pv_{\!\!\!\! 0}^L
              dq'\,{{\delqp\eta}\over{x(q)-x(q')}}\right)^2  \cr
   &= -{\pi L^3\over 24} \sum_{mnp>0}{}^{\!\!\!\!\!
       \prime}\,\phi_m\phi_n\phi_p\,
         \bigl(-3|k_m+k_n+k_p|+|k_m-k_p-k_n|  \cr
   &\quad\, +|k_m+k_p-k_n|
             +|k_m-k_p+k_n|\bigr),
}   \eqn\naiveExpansion
$$
where the result \kernelB\ has been used.  Starting from the right
hand side of \qspaceId\ as in the bosonic collective field theory,
one finds
$$
\eqalign{
   {\pi^2\over 3}\int &{dq\over\phi_0^2}\,(\delq\eta)^3   \cr
   &= -{\pi L^3\over 24} \sum_{mnp>0}{}^{\!\!\!\!\!
        \prime}\,\phi_m\phi_n\phi_p\,
         \bigl(|k_m+k_n+k_p|+|k_m-k_p-k_n|    \cr
   &\quad\, +       |k_m+k_p-k_n|
          +|k_m-k_p+k_n|\bigr),           \cr
}   \eqn\cftExpansion
$$
which is manifestly different from the left hand side.

It is therefore evident that
the regularization scheme of chapter 5 differs from the one used in
the bosonic collective field theory.
The reason for the difference between \naiveExpansion\ and
\cftExpansion\ is as follows:  In [20] the turning point divergence
in \cftExpansion\ was regulated essentially by introducing a cutoff
$\epsilon$ so that the integration range becomes
$\int_\epsilon^{L-\epsilon}$.  Keeping epsilon small but finite
throughout and
only at the end discarding all $\epsilon$-dependent quantities (which
are argued to be nonuniversal), one obtains the principal part
prescription.   Applying the same cutoff to \naiveExpansion\
corresponds to identifying the $\epsilon$-independent terms in
$$
\int_\epsilon^{L-\epsilon}dq\,\dots\left(\pv_{\!\!\!\!
\epsilon}^{L-\epsilon}
      dq'\,\dots
   \right)^2.  \eqno\eq
$$
As it stands, in \naiveExpansion\ we calculated not the quantities \?
but instead
$$
\int_\epsilon^{L-\epsilon}dq\,\dots\left(\pv_{\!\!\!\! 0}^{L}
      dq'\,\dots
   \right)^2.  \eqno\eq
$$

If we want correspondence with the bosonic cubic collective field
theory, we must therefore subtract
corrections of the typical form
$$
\int_\epsilon^{L-\epsilon}dq\,\dots\left(\pv_{\!\!\!\! 0}^{L}
      dq'\,\dots
   \right)
    \left(\int_{0}^{\epsilon}
      dq''\,\dots
   \right)  \eqno\eq
$$
from \naiveExpansion.

Explicitly, using \contourId\ and remembering that $x \propto \cos \pi
q/L$ and $\phi_0 \propto \sin \pi q/L$, we have
$$
\eqalign{
   \int_\epsilon^{L-\epsilon} & dq\,\cos k_m q \,\pv_{\!\!\!\! 0}^{L}
      dq'\,{\cos k_n q'\over x(q) - x(q')}
   \, \int_{0}^{\epsilon}
      dq''\,{\cos k_p q''\over x(q)-x(q'')}
       \cr
  &\propto
   \int_\epsilon^{L-\epsilon} dq\,{\cos k_m q \sin k_n q \over
      \phi_0(q)}
    \,\int_{0}^{\epsilon}
      dq''\,{\cos k_p q''\over x(q)-x(q'')}
       \cr
  &\sim
   \int_\epsilon^{L-\epsilon}  dq\,{\cos k_m q \sin k_n q \over
      \sin {\pi q\over L}}
    \epsilon\left(
      {1\over \cos {\pi q\over L}-1}
   \right).    \cr
}  \eqn\approx
$$
To see how we obtain a finite contribution from this, note that one
gets
an $o(\epsilon)$ contribution from the integral $\int_0^\epsilon dq''$
and an $o(1/\epsilon)$ contribution (see below) from the lower limit
of the integral $\int_\epsilon^{L-\epsilon} dq$.  The
$\epsilon$-dependence will therefore cancel.  More precisely, the
small $q$ behaviour of the integrand in \approx\ is given by
$$
  {\cos k_m q}\,{\sin {n\pi q\over L}\over\sin{\pi q\over L}}\,
      {1\over \cos {\pi q\over L}-1}
  \sim {n\over \half {\pi^2 q^2\over L^2}}
  \propto {k_n\over q^2}.
$$
Inserting this into the integral in \approx, one gets contributions of
the form
$$
  \epsilon \int_\epsilon dq\,{k_n\over q^2}
  \sim \epsilon\, k_n\, {1\over\epsilon} \sim k_n.
$$

We therefore see that these subtractions give rise to finite (\ie,
$\epsilon$-independent) corrections to the expansion \naiveExpansion\
proportional to
$$
  \sum_{mnp>0} k_n\,\phi_m\phi_n\phi_p =
  {1\over 3}\sum_{mnp>0} (k_m +k_n+k_p)\,\phi_m\phi_n\phi_p.
   \eqno\eq
$$
These corrections are exactly of the right form and the right sign to
restore equality between \naiveExpansion\ and \cftExpansion.  There
are two such terms, corresponding to the lower limit of
the integrations over each of $q'$ and $q''$.  Due to the crudeness
of the approximation in \approx, the above argument is not expected
to give the correct overall coefficient of the correction term.
However, this coefficient can be inferred from the fact that for a
finite cutoff $\epsilon$, all integrals are well behaved and the
identity \qspaceId\ will be valid.  We therefore expect the correction
term to be exactly equal to the difference between \naiveExpansion\
and \cftExpansion, which is given by
$$
   -{\pi L^3\over 24} \sum_{mnp>0}{}^{\!\!\!\!\!
       \prime}\,4\,(k_m+k_n+k_p)\,\phi_m\phi_n\phi_p.
$$

Once the coefficients have been determined, we can can work backwards
and write down the {\it regulated\/} expressions
$$
\eqalign{
  \int dq\, {\cos k_m q \sin k_n q \over \phi_0^2}
     &\left(\phi_0 \pv dq'' {\cos k_p q''\over x(q) - x(q'')}\right)
   \cr
  &= - {L^3\over 8}\bigl(|k_m+k_n-k_p| + |k_m-k_n+k_p|
     \cr
  &\quad -|k_m+k_n+k_p| - |k_m-k_n-k_p| + 2k_n    \bigr)
}   \eqn\cubicQref
$$
and
$$
\eqalign{
  \int dq\, {\cos k_m q \over \phi_0^2}
     &\left(\phi_0 \pv dq' {\cos k_n q'\over x(q) - x(q')}\right)
     \left(\phi_0 \pv dq'' {\cos k_p q''\over x(q) - x(q'')}\right)
   \cr
  &= - {\pi L^3\over 8}\bigl(|k_m+k_n-k_p| + |k_m-k_n+k_p|
     \cr
  &\quad -|k_m+k_n+k_p| - |k_m-k_n-k_p| + 2k_n + 2k_p   \bigr),
}   \eqno\eq
$$
where the final terms are cutoff independent corrections to
the naive expressions obtained from the principal part prescription.

Using the second of these two equalities in the calculation of the
expansion of
\naiveExpansion, we indeed find equivalence between \naiveExpansion\
and \cftExpansion.

Equation \cubicQref\ can be used in the calculation
of the cubic part of the supercharge $Q$ of the previous section.
However this regularization is
{\it inconsistent\/} with supersymmetry in the following sense:
because the quadratic part of the supercharge receives no
turning point correction,  if one evaluates
$H=\half\{Q,Q^\dagger\}$, one finds coefficients which are
inconsistent with \?.
In other words, the regularization does not commute with taking the
bracket.

This is not unrelated to what is happening in the continuum
approach to $d=1$ superstrings [6].  If one uses the identity
\xspaceId\ in $x$-space first and then changes variables to $q$-space,
in the process rendering the interactions local, one is not able, at
least in our approach, to generate this term from the supercharges.
This does not rule out the possibility that one may find a local
supersymmetric
extension of the bosonic collective string theory in $x$-space (this
can certainly be done at the quadratic level [23]).

It should be emphasized if the integrals are regulated
as described in chapter 5, the theory is fully supersymmetric.

\chapter {CONCLUSIONS}

\noindent
In this paper an investigation was made of the Marinari-Parisi model
in an arbitrary potential with particular emphasis on its continuum
formulation and possible
application in the context of supersymmetrizing the bosonic collective
$d=1$ string field theory.

We started by indicating the equivalence of the Marinari-Parisi
model, the
super-Calogero system and the supersymmetrization of the collective
field theory defined in [14].
Developing the perturbation theory, we saw that,
as a result of nontrivial commutation relations amongst
continuum fields,
the hamiltonian acquired an infinite sequence of vertices, in contrast
to the cubic bosonic collective field theory where only a
cubic vertex was required.

Assuming that supersymmetry is preserved at the level of the leading
semiclassical
configuration, we demonstrated
supersymmetry of the semiclassical spectrum and,
for a specific class
of potentials,
we showed that the spectrum consists of a massless boson and
Majorana fermion.

An investigation of the scaling behaviour of the theory showed that
for potentials which are
homogeneous of arbitrary degree in the cosmological constant one can
identify a
rescaling of variables in terms of which the cosmological constant is
scaled out while the normalization condition remains the same.
This implies that in general
the time of flight is finite and
that the mechanism of generation of an infinite Liouville-like
dimension [13, 22] is not present in this case.
The spacetime
interpretation of the Marinari-Parisi model therefore remains
problematic and more work is needed on this issue.

An examination of the bosonic sector cubic oscillator vertices in the
harmonic case showed a discrepancy with
those of the collective bosonic string field theory [20]. This
discrepancy resulted from a difference between the turning
point regularization prescriptions of chapter 5 and reference [20].
Adjusting the regularization to agree with that of [20], we found that
a bosonic sector compatible with the bosonic collective
field theory cannot be generated from a corresponding
regularization of the supercharges.  In other words, the
regularization does not commute with the bracket $H = \{Q,
Q^\dagger\}$.

We have indicated that, provided a local supersymmetrization of the
bosonic collective field theory can be found, it is likely that this
problem will be overcome.  Probably the best context in which this can
be attempted is the potential free case [23].  In the bosonic case,
this
corresponds to the background independent formulation of the theory,
and a
W-algebra structure has been identified [18, 32].
The presence of this large symmetry is related to the existence of an
infinite number of conserved quantities in the underlying Calogero
model.
In the super-Calogero model with a harmonic
oscillator potential
a few low lying exact states have already been constructed [15].
An extension of this analysis may provide
a link to a suitable supersymmetric
generalization of the W-algebra [33].

\bigskip

\noindent
{\bf ACKNOWLEDGEMENTS:}
We wish to thank Antal Jevicki for some stimulating discussions and
Kre\v simir Demeterfi for his very careful reading of the manuscript.
We are grateful towards the theory group at Brown University for their
hospitality during our stay.

\vfill
\endpage

\centerline{\bf REFERENCES}
\bigskip
\pointbegin
    D. J. Gross and A. A. Migdal, {\it Phys. Rev. Lett.} {\bf 64}
(1990)
    127;
      M. R. Douglas and S. H. Shenker, {\it Nucl. Phys.} {\bf B335}
    (1990) 635;
      E. Br\'ezin and V. A. Kazakov, {\it Phys. Lett.} {\bf B236}
    (1990)
    144;
       C. Crnkovic, P. Ginsparg and G. Moore, {\it Phys. Lett.}
    {\bf B237} (1990) 196;
      D. J. Gross and A. A. Migdal, {\it Phys. Rev. Lett.} {\bf 64}
    (1990) 717;
      E. Br\'ezin, M. Douglas, V. A. Kazakov and S. Shenker, {\it
    Phys.
    Lett.} {\bf B237}
      (1990) 43; M. Douglas, {\it Phys. Lett.} {\bf B238} (1990) 176.
\point
    D. J. Gross and N. Miljkovic, {\it Phys. Lett.} {\bf B238} (1990)
        217;
        E. Br\'ezin, V. A. Kazakov and A. B. Zamolodchikov,
        {\it Nucl. Phys.} {\bf B338} (1990) 673;
        P. Ginsparg and J. Zinn-Justin, {\it Phys. Lett.} {\bf B240}
        (1990) 333.
\point
    I. R. Klebanov,
    Princeton preprint PUPT-1271 (1991).
\point
    A. D'Adda, {\it Class. Quant. Grav.} {\bf 9} (1992) L21;
    L. Alvarez-Gaum\'e and J. L. Ma\~nes,
    {\it Mod. Phys. Lett.} {\bf A6} (1991) 2039.
\point
    E. Abdalla, M. C. B. Abdalla, D. Dalmazi and  K. Haradi,
        S\~ao Paolo preprint PRINT-91-0351 (1991);
        K. Aoki and E. d'Hoker, {\it Mod. Phys. Lett.} {\bf A7} (1992)
        333;
        L. Alvarez-Gaum\'e and P. Zaugg,
        {\it Phys. Lett.} {\bf B273} (1991) 81.
\point
    P. Di Francesco and D. Kutasov,
    Princeton University preprint
        PUPT-1276 (1991).
\point
    P. Bouwknegt, J. McCarthy and K. Pilch,
    CERN preprint CERN-TH.6346/91 (1991);
    K. Itoh and N. Ohta,
    Osaka University and Brown University
    preprints  OS-GE 22-91, BROWN-HET-844 (1992).
\point
        E. Marinari and G. Parisi, {\it Phys. Lett.} {\bf B240} (1990)
        375.
\point
    E. Br\'ezin, Talk delivered at the XXVth International Conference
    on
    High Energy Physics, 1990.
\point
     J. Feinberg,
     Israel Inst. of Technology
     preprint TECHNION-PH-92-1 (1992).
\point
    J. Alfaro and P. H. Damgaard, {\it Phys. Lett.} {\bf B222} (1989)
        425;
        M. Karliner and A. A. Migdal, {\it Mod. Phys. Lett.} {\bf A6}
        (1990) 2565;
        S. Belluci, T. R. Govindrajan, A. Kumar and R. N. Oerter,
        {\it Phys. Lett.} {\bf B249} (1990) 49;
        J. Gonzalez, {\it Phys. Lett.} {\bf B255} (1991) 367.
\point
    A. Jevicki and B. Sakita, {\it Nucl. Phys.} {\bf B165} (1980) 511.
\point
    S. R. Das and A. Jevicki, {\it Mod. Phys. Lett.} {\bf A5} (1990)
    1639.
\point
    A. Jevicki and J. P. Rodrigues, {\it Phys. Lett.} {\bf B268}
(1991)
    53.
\point
    D. Z. Freedman and P. F. Mende, {\it Nucl. Phys.} {\bf B344}
(1990)
    317.
\point
  F. Calogero, {\it J. Math. Phys.} {\bf 10} (1969) 2191,
  {\it J. Math. Phys.} {\bf 10} (1969) 2197,
  {\it J. Math. Phys.} {\bf 12} (1971) 419,
  {\it Lett. Nuovo Cim.} {\bf 19} (1977) 505,
  {\it Lett. Nuovo Cim.} {\bf 20} (1977) 251;
  A. M. Perelomov, {\it Teor. Mat. Fiz.} {\bf 12} (1971) 364,
  {\it Ann. Inst. Henri Poincar\'e} {\bf 28} (1978) 407;
  M. A. Olshanetsky and A. M. Perelomov, {\it Phys. Rep.} {\bf 71}
  (1981) 313, {\it Phys. Rep.} {\bf 94} (1983) 313;
  M. Bruschi and F. Calogero, {\it Lett. Nuovo Cim.} {\bf 24} (1879)
  601;
  S. Ahmed, M. Bruschi, F. Calogero, M. Olshanetsky and A. M.
  Perelomov, {\it Nuovo Cim.} {\bf B49} (1979) 173.
\point
    A. Jevicki and H. Levine, {\it Phys. Rev. Lett.} {\bf 44} (1980)
    1443.
\point
    J. Avan and A. Jevicki, {\it Phys. Lett.} {\bf B266} (1991) 35;
    {\it Phys. Lett.} {\bf B272} (1991) 17;
    Brown University preprint
       BROWN-HET-839
       (1991).
\point
        A. Dabholkar, Rutgers University preprint RU-91-20 (1991).
\point
    K. Demeterfi, A. Jevicki and J. P. Rodrigues,
        {\it Nucl. Phys.} {\bf B362} (1991) 173;
        {\it Nucl. Phys.} {\bf B365} (1991) 499;
    {\it Mod. Phys. Lett.} {\bf A6} (1991) 3199.
\point
     J. P. Rodrigues, Ph.D. thesis, Brown University (1983);
     A. Jevicki and J. P. Rodrigues, {\it Nucl. Phys.} {\bf
     B230} [FS10] (1984) 317.
\point
    J. Polchinski, {\it Nucl. Phys.} {\bf B346} (1990) 253.
\point
    A. J. van Tonder, Ph.D. Thesis, University of the Witwatersrand
   (1992); University of the Witwatersrand preprint CNLS-92-01 (1992).
\point
   J. D. Cohn and H. Dykstra,
   FERMILAB-PUB-92/35-T (1992).
 \point
    P. Di Francesco and D. Kutasov, {\it Phys. Lett.} {\bf B261}
      (1991)
        385;
        N. Sakai and Y. Tanii,
        {\it Int. J. Mod. Phys.} {\bf A6} (1991) 2743;
        Y. Kitazawa, Harvard preprint HUTP-91/A034 (1991);
        M. Bershadsky and I. R. Klebanov, {\it Phys. Rev. Lett.} {\bf
        65} (1990) 3088;
        {\it Nucl. Phys.} {\bf B360} (1991) 559;
\point
    A. Jevicki,
    Brown University preprint BROWN-HET-807 (1991).
\point
    E. Br\'ezin, C. Itzykson, G. Parisi and J. B. Zuber, {\it Commun.
    Math. Phys.} {\bf 59}, 35-51 (1978).
\point
    M. B. Green, J. H. Schwarz and E. Witten, \lq\lq Superstring
    Theory\rq\rq (Cambridge University Press, 1987) Volume I.
\point
   P. A. M. Dirac, \lq\lq Lectures on Quantum Mechanics\rq\rq,
   (Yeshiva University press, 1964).
\point
   F. A. Berezin, \lq\lq The Method of Second Quantization\rq\rq,
   (Academic Press, 1966).
\point
    E. Witten, {\it Nucl. Phys.} {\bf B185} (1981) 513,
    {\it Nucl. Phys.} {\bf B202} (1982) 253;
    P. Salomonson and J. W. van Holten, {\it Nucl. Phys.} {\bf B196}
       (1982) 509;
    D. Lancaster, {\it Nuovo Cim.} {\bf 79A} (1984) 28.
\point
    J. Polchinski, {\it Nucl. Phys.} {\bf B362} (1991) 125;
    D. Minic, J. Polchinski and Z. Yang,
    {\it Nucl. Phys.} {\bf B369} (1992) 324;
    E. Witten,
     IASSNS-HEP-91/51
     (1991), to appear in {\it Nucl. Phys.} {\bf B};
    E. Witten and B. Zwiebach,
    SLAC-PUB-IASSNS-HEP-92/4, MIT-CTP-2057 (1992).
\point
  E. Bergshoeff, B. de Wit and M. Vasiliev,
  CERN preprint CERN-TH.6021/91;
  E. Bergshoeff, C. N. Pope, L. J. Romans, E. Sezgin and X. Shen,
  {\it Phys. Lett.} {\bf B245} (1990) 447;
  T. Inami, Y. Matsuo and I. Yamanaka, {\it Phys. Lett.}
  {\bf B215} (1988) 701;
  F. Yu,
  Utah University preprint
  UU-HEP-91/12 (1991).
\point
  B. Zwiebach, {\it Ann. Phys.} {\bf 186} (1988) 111.
\point
  J. Cohn and S. P. DeAlwis, {\it Nucl. Phys.} {\bf B368} (1992) 79.
\point
   K. Demeterfi, N. Deo, S. Jain
   and Chung-I Tan, {\it Phys. Rev.} {\bf D42} (1990) 4105.

\vfill
\endpage

\end